\documentclass{aastex6}

\usepackage{multirow}
\usepackage{subfigure}

\begin{document}
\title{A new 95 GHz methanol maser catalog: I. data}

\author{Wenjin Yang\altaffilmark{1,3,4}, Ye Xu\altaffilmark{1}, Xi Chen\altaffilmark{2,5}, Simon P. Ellingsen\altaffilmark{6}, Dengrong Lu\altaffilmark{1}, Binggang Ju\altaffilmark{1}, Yingjie Li\altaffilmark{1,2}}

\altaffiltext{1}{Purple Mountain Observatory, Chinese Academy of Science, Nanjing 210008, China; wjyang@pmo.ac.cn}
\altaffiltext{2}{Center for Astrophysics, GuangZhou University, Guangzhou 510006, China; xuye@pmo.ac.cn}
\altaffiltext{3}{University of Science and Technology of China, 96 Jinzhai Road, Hefei 230026, China; chenxi@shao.ac.cn}
\altaffiltext{4}{Graduate University of the Chinese Academy of Sciences, 19A Yuquan Road, Shijingshan District,
Beijing 100049, China}
\altaffiltext{5}{Shanghai Astronomical Observatory, Chinese Academy of Sciences, Shanghai 200030, China.}
\altaffiltext{6}{School of Physical Sciences, University of Tasmania, Hobart, Tasmania, Australia}

\begin{abstract}

The Purple Mountain Observatory 13.7 m radio telescope has been used to search for 95 GHz (8$_0$--7$_1$A$^+$) class I methanol masers towards 1020 Bolocam Galactic Plane Survey (BGPS) sources, leading to 213 detections. 
We have compared the line width of the methanol and HCO$^+$ thermal emission in all of the methanol detections and on that basis we find 205 of the 213 detections are very likely to be masers.  This corresponds to an overall detection rate of 95 GHz methanol masers towards our BGPS sample of 20\%. Of the 205 detected masers 144 (70\%) are new discoveries. Combining our results with those of previous 95 GHz methanol masers searches, a total of four hundred and eighty-one 95 GHz methanol masers are now known, we have compiled a catalog listing the locations and properties of all known 95 GHz methanol masers.

\end{abstract}


\section{Introduction} \label{sec:intro} 
Methanol masers are classified into two types based on their association with other star-forming phenomena \citep{1991ASPC...16..119M}. Class~II methanol masers, such as the 6.7 GHz 5$_{1}$--6$_{0}$A$^{+}$ and 12.2 GHz 2$_{0}$--3$_{-1}$E transitions \citep{1991ApJ...380L..75M,1987Natur.326...49B}, are often found in close proximity to ultracompact H{\scriptsize II} regions, infrared sources and OH masers \citep{1998MNRAS.301..640W}, and are known to be radiatively pumped \citep{1992MNRAS.259..203C,2005MNRAS.360..533C}. Whereas class I methanol masers are often found offset by $\sim$1$\arcmin$ ($\sim$0.1--1pc) from the other maser species, infrared sources etc, the collisional pumping of the class I methanol masers is not only suggested by the association with astrophysical shocks but also supported by theoretical predictions \citep{1992MNRAS.259..203C,1996IAUS..178..163M,2014ApJ...793..133M,2016A&A...592A..31L}.

Compared to class~II methanol masers, class~I methanol masers are relatively poorly studied and understood. One reason for this is that the strongest class I methanol maser transitions occur at relative high frequencies, such as the 44 GHz 7$_{0}$--6$_{1}$A$^{+}$ and 95 GHz 8$_{0}$--7$_{1}$A$^{+}$ transitions. In addition, a lack of target sources which are expected to be associated with class I methanol masers has increased the difficulty of searching for new maser sites. Nevertheless, searches for class I methanol masers have advanced significantly in recent years due to the improved performance of single dish telescopes at millimeter wavelengths and the greater availability of data at infrared and longer wavelengths from surveys of the Galactic plane with high sensitivity and angular resolution. Recent interferometric observations \citep{2016ApJS..222...18G,2014MNRAS.439.2584V,2010ApJS..191..207G,2004ApJS..155..149K} of 44 GHz class I methanol masers have improved our understanding of this transition. For the 95 GHz methanol masers, surveys towards the $Spitzer$ Galactic Legacy Infrared Mid-Plane Survey Extraordinaire (GLIMPSE) Extended Green Objects (EGOs) \citep{2011ApJS..196....9C,2013ApJS..206....9C}, molecular outflow sources \citep{2013ApJ...763....2G} and cross-matching between the GLIMPSE point sources and the Bolocam Galactic Plane Survey (BGPS) sources \citep{2012ApJS..200....5C} have significantly increased the number of 95 GHz methanol masers that are known and the number of known class I maser regions in general. To date, high-resolution observations of the 95 GHz transition have only been made towards a few famous star-forming regions (e.g., Orion--KL \citep{1988ApJ...330L..61P}, DR 21 \citep{1990ApJ...364..555P}, W 33 and W 51 \citep{1993LNP...412..211P}). The expectation is that the distribution of the 95 GHz transition will be similar to that of the 44 GHz masers, however, further interferometric observations are required to confirm this and to determine the detailed relationship between the different class I methanol maser transitions.

There are several reasons as to why investigation of 95 GHz methanol masers is required: The 95 GHz transition is one of the strongest and most commonly detected class I methanol maser transitions, which may provide a unique tool for investigating outflows activity where other standard shock tracers are too weak to be detected \citep{2016A&A...592A..31L}. 
Within the maser-based evolutionary timeline for high-mass star formation \citep{2010MNRAS.401.2219B}, the relationship of class I methanol masers to other types of masers is still unclear.  95 GHz methanol masers are potentially a very useful tool for investigating high-mass star formation and inclusion of class~I masers within this timeline remains an outstanding issue. 
Moreover, where different class I methanol maser transitions are spatially coincident studies of the 95 GHz masers will help investigation of the pumping mechanism.

In this paper, we report a search for 95 GHz methanol masers towards BGPS sources which has been undertaken using the Purple Mountain Observatory (PMO) 13.7 m radio telescope in Qinghai province, China. In Section \ref{sec:selobs} we describe the source selection and observations, in Section \ref{sec:results} we present the results of this new survey and we have also compiled a catalog of all known 95 GHz methanol masers (Section \ref{sec:catalog}), followed by summary in Section \ref{sec:summary}.
In an upcoming publication (paper II), we will focus on statistical analysis of the 95 GHz methanol maser sources using molecular spectroscopy, infrared and radio continuum data to better understand the molecular environments.

\section{Source Selection and Observations} \label{sec:selobs}

\subsection{Source Selection} \label{subsec:source selection}
Our sample is selected from the BGPS\footnote{\url{http://irsa.ipac.caltech.edu/data/BOLOCAM_GPS/}} \citep[version 1.0.1][]{2010ApJS..188..123R,2011ApJS..192....4A}, which is a 1.1 mm continuum survey of the galactic plane in the northern hemisphere undertaken using the Caltech Submillimeter Observatory. 
\citet{2011ApJS..196....9C} searched for class I methanol masers towards a sample of high-mass star formation regions with outflows (EGOs) identified from the {\em Spitzer} GLIMPSE survey and found that the detection rate was higher towards those that had an associated BGPS source.

From 8358 BGPS sources, we selected 1020 sources ($ > $ 10\% of the total BGPS sources) in total.
There are 982 BGPS sources which meet the criteria identified by \citet{2012ApJS..200....5C} for BGPS sources likely to have an associated class I methanol maser.
The criteria were $ \rm{\log}$($S_{\rm int}$)$ \leq -38.0+1.72\;\rm{\log}$($N^{\rm beam}_{\rm H_{2}})$, $ \rm \log$($N^{\rm beam}_{\rm H_{2}}$) $\geq 22.1$, where $N^{\rm beam}_{\rm H_{2}}$ and $S_{\rm int}$ are the beam-averaged column density and integrated flux density, respectively from the BGPS point source catalog.  These criteria define a region of the beam-averaged column density -- integrated flux density plane which includes 90\% of the 95 GHz methanol masers detected in a cross-matched sample of GLIMPSE point sources and BGPS clumps.   

Recently a new method of searching for class I methanol masers was proposed by \citet{2016PASA...33...15M}. 
They applied three different statistical classification techniques to three different size astronomical datasets to compare the performance of these methods in identifying whether specific mid-infrared or millimeter continuum sources are likely to be associated with interstellar masers.
The three statistical classification techniques investigated were linear discriminant analysis (LDA), logistic regression and random forests.
LDA and logistic regression are parametric methods, the former uses the linear combination of predictor variables that maximizes the separation of the different classes and minimizes the variation within classes to convert high-dimensional data to a real number, then the classification of this sample is determined by comparing this number to a threshold number \citep{2012msma.book.....F}. The latter is a form of generalized linear modeling that is used to predict the probability of an event occurring, and maximum likelihood is used to estimate the parameters of the model. 
Random forests is a non-parametric method which uses multiple classification trees produced from training data, and classifies based on the results of the majority of the classification trees for a particular input.
 \citeauthor{2016PASA...33...15M} found that for small data sets parametric methods perform better than random forests, while for larger data sets random forests surpass the others.
All the statistical classification techniques are suited for their specific goal of identifying BGPS sources which are more likely to have an associated class I methanol masers.

The statistical classification techniques were trained using the results of the 95 GHz methanol maser search of \cite{2012ApJS..200....5C} which was targeted towards regions selected on the basis of their mid-infrared and millimeter wavelength emission. 
The mid-infrared data was taken from the {\em Spitzer} GLIMPSE survey, while the millimeter continuum data was from the BGPS. 
The training dataset consisted of 214 BGPS sources of which 62 were associated with 95 GHz class I methanol masers and the remainder were not. Each of the statistical classification techniques were trained using these results and then applied to the remaining 8144 BGPS sources which were not part of the training dataset.
The predictor variables used in the classification models were the angular size of the major and minor axis of the dust clump, position angle, deconvolved angular radius, 1.1-mm flux density within apertures of diameter 40, 80, and 120 arcsec and the integrated flux density.
From the sample of 8144 BGPS sources a total of 739 were identified as having a greater than 50\% chance of having an associated 95 GHz class I methanol maser by one or more of the three statistical classification techniques.
The majority of the 739 \citeauthor{2016PASA...33...15M} predictions were included in the initial sample of 982 BGPS sources.  For completeness we also made observations of the remaining 38 Manning et al. sources that were not.

In addition, all the sources must have declination greater than --25$^{\circ}$ (so as to be accessible with the PMO 13.7 m radio telescope), and the separation of each target source from all other the sources in the sample is greater than the beam size of the PMO 13.7 m telescope at 95 GHz ($\sim$1$\arcmin$)(where this was not the case the BGPS source with the greater value of $N^{\rm beam}_{\rm H_{2}}$ was chosen as the target). Moreover, for all of the sources in our sample the telescope was pointed at the position of the 1.1 mm BGPS source peak.

\subsection{Observations} \label{subsec:obs}

The observations of the 8$_{0}$--7$_{1}$A$^{+}$ class I methanol maser transition (assumed rest frequency 95.1694630 GHz) \citep{2004A&A...428.1019M} and 89.188526 GHz HCO$^+$ ($J$=1--0) were made with the PMO 13.7 m radio telescope in Delingha, China. Of the total sample of 1020 BGPS sources, the 982 BGPS sources which meet the criteria identified by \citet{2012ApJS..200....5C} were observed in the period May--June each year between 2012 and 2015. And the 38 BGPS sources selected from \cite{2016PASA...33...15M} were observed in 2016 June. A cryogenically cooled nine-beam superconductor-insulator-superconductor (SIS) receiver was used to observe the 95 GHz methanol maser and 89 GHz HCO$^+$ transitions simultaneously in the upper and lower sideband, respectively. This receiver operates over a frequency range of 85--115 GHz and the central beam of the nine-beam receiver was pointed at the target position in most cases. The spectra were recorded using a Fast Fourier Transform Spectrometer with 16384 spectral channels across a bandwidth of 1 GHz (corresponding to a velocity range of about 3000 km s$^{-1}$) with a frequency resolution of 61 kHz.  This yields an effective velocity resolution of 0.19 km s$^{-1}$ and 0.21 km s$^{-1}$ for 95 GHz methanol masers and HCO$^+$, respectively. We searched the velocity range from --200 km s$^{-1}$ to 200 km$^{-1}$ (local standard of rest frame), because this covers the expected velocity of any class I maser emission in the Milky Way. The system temperature for the 95 GHz methanol maser observations was in the range 135--325 K, depending on the weather conditions and telescope elevation. Most sources were observed in a position-switching mode with an off-position offset of 10$\arcmin$ in right ascension.  For some sources a different reference position was chosen to ensure no emission in the reference spectrum. The pointing rms was better than 5$\arcsec$. The standard chopper wheel calibration technique \citep{1976ApJS...30..247U} was applied to measure an antenna temperature, $T^{*}_{\rm A}$, corrected for atmospheric absorption. The beam size of the telescope is approximately 55$\arcsec$ at 95 GHz, with a main beam efficiency $\eta_{\rm mb}$ of 59\%. The antenna efficiency is 47\%, corresponding a factor of 39.9 Jy K$^{-1}$ to convert antenna temperature into flux density. The system temperature for the HCO$^+$ observations varied between 130--300 K. The beam size is about 1$\arcmin$ at 89 GHz, with a main beam efficiency $\eta_{\rm mb}$ of 57\%. The antenna efficiency is 46\%, corresponding a factor of 40.7 Jy K$^{-1}$ to convert antenna temperature into flux density. The initial search used an on-source integration time of 10 minutes for each source achieving rms noise levels of approximately 1.2 Jy and 1.4 Jy for the 95 GHz methanol masers and HCO$^+$, respectively. Then, depending on whether any emission was detected and its intensity, we observed for a further 10--30 minutes to improve the signal-to-noise ratio of the final spectrum, reaching typical rms noise levels of 0.7 Jy and 0.8 Jy for 95 GHz methanol and HCO$^+$, respectively. 

The data were reduced and analyzed using the GILDAS/CLASS package. Although data from all nine beams were recorded, we only analyzed the data for the one beam used to tracked the target position, as the other eight beams rotated during the observation.  A low-order polynomial baseline subtraction and Hanning smooth were performed on all averaged spectra. The majority of the 95 GHz methanol detections do not show a simple Gaussian profile, this is likely because many of the spectra contain multiple maser features with overlapping velocities. Nevertheless, to characterize the spectra, we undertook Gaussian fitting of each peak in the spectrum for each source. The data reduction procedure for the 89 GHz HCO$^+$ thermal emission was the same as that used for the methanol masers.

\section{Results} \label{sec:results}

\subsection{95 GHz Methanol Masers Detection} \label{subsec:detection}
 
\begin{figure}
\mbox{
\begin{minipage}[b]{4.1cm}
\includegraphics[scale=0.23,angle=-90]{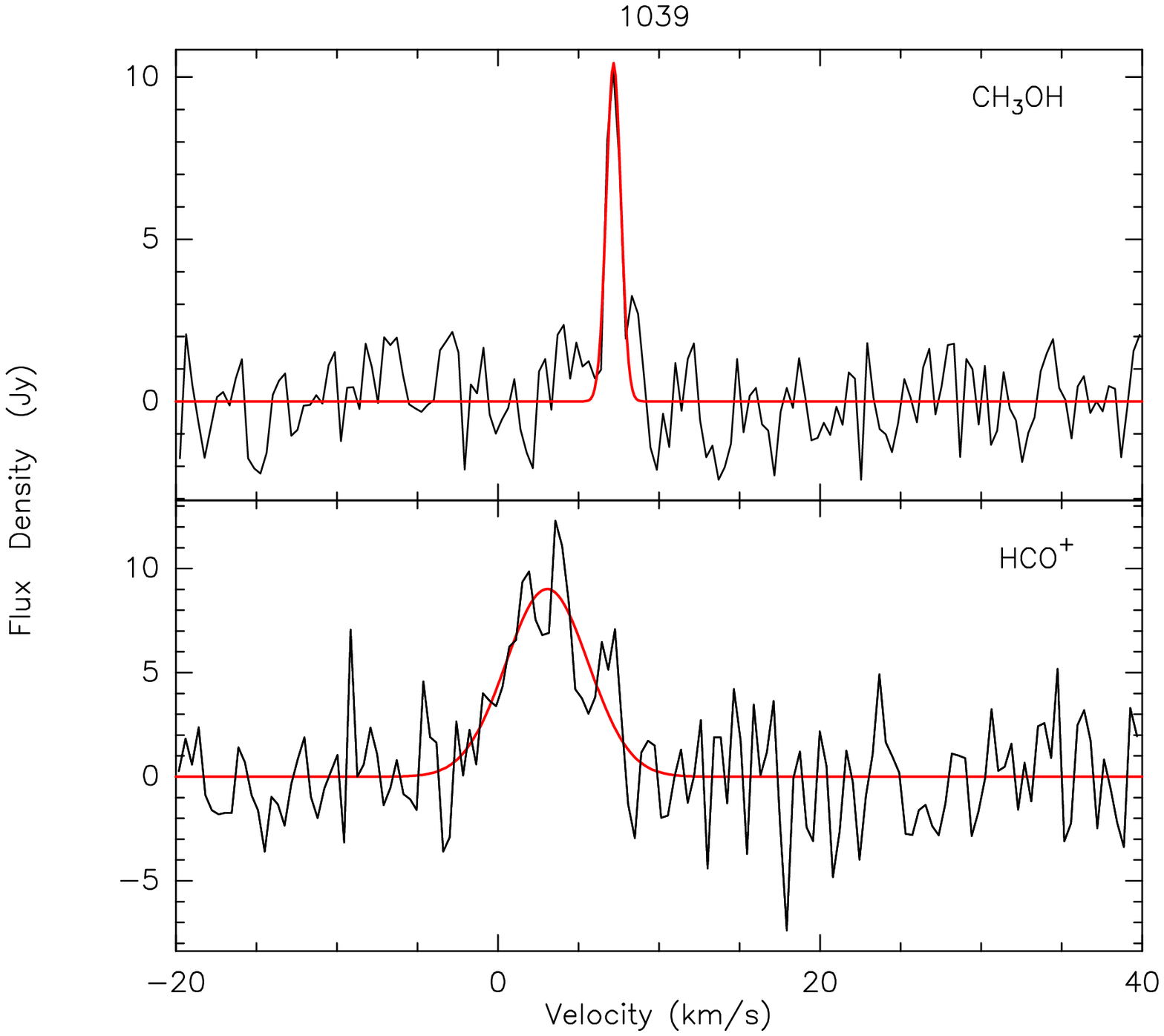}
\end{minipage}
\begin{minipage}[b]{4.1cm}
\includegraphics[scale=0.23,angle=-90]{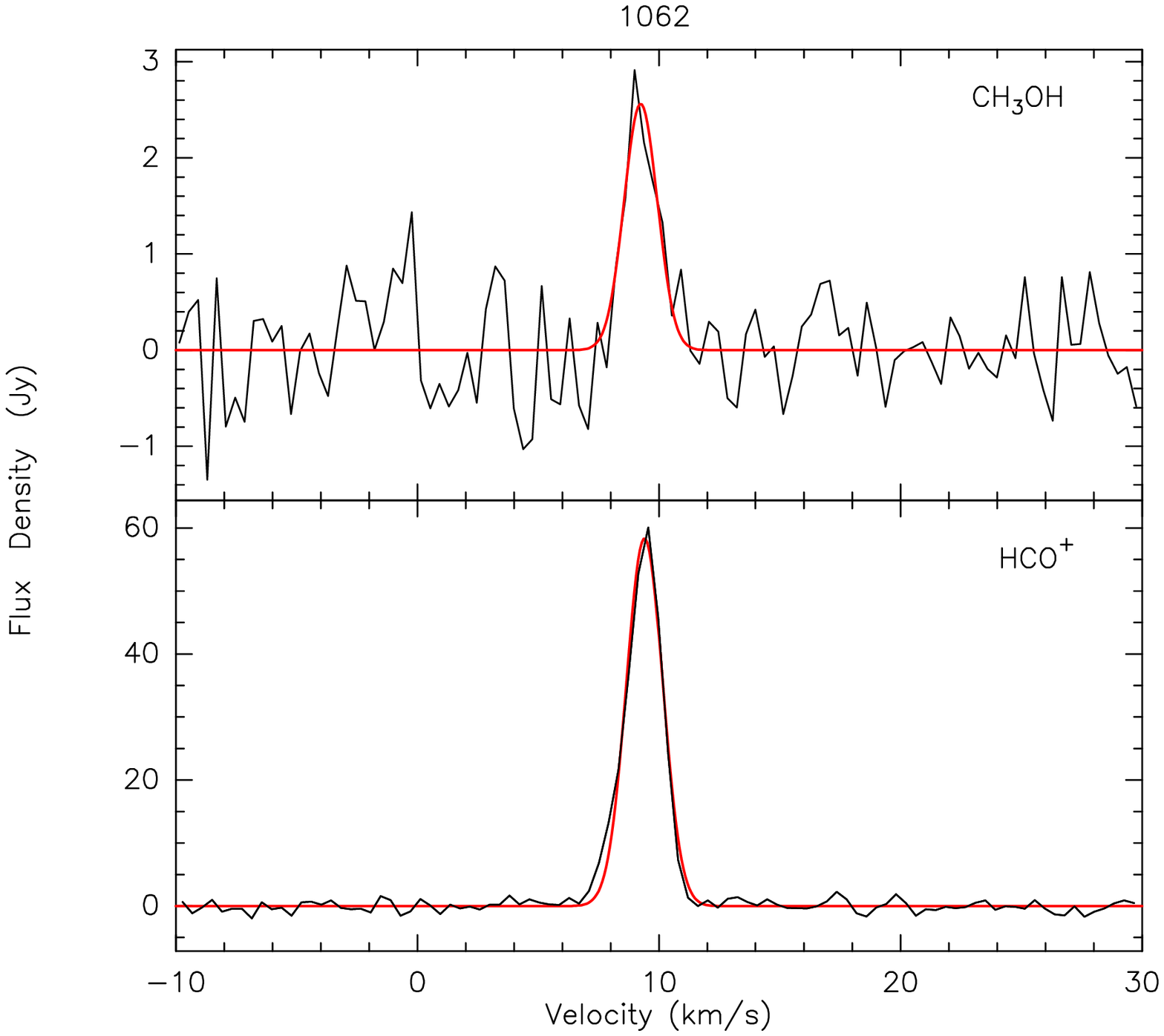}
\end{minipage}
\begin{minipage}[b]{4.1cm}
\includegraphics[scale=0.23,angle=-90]{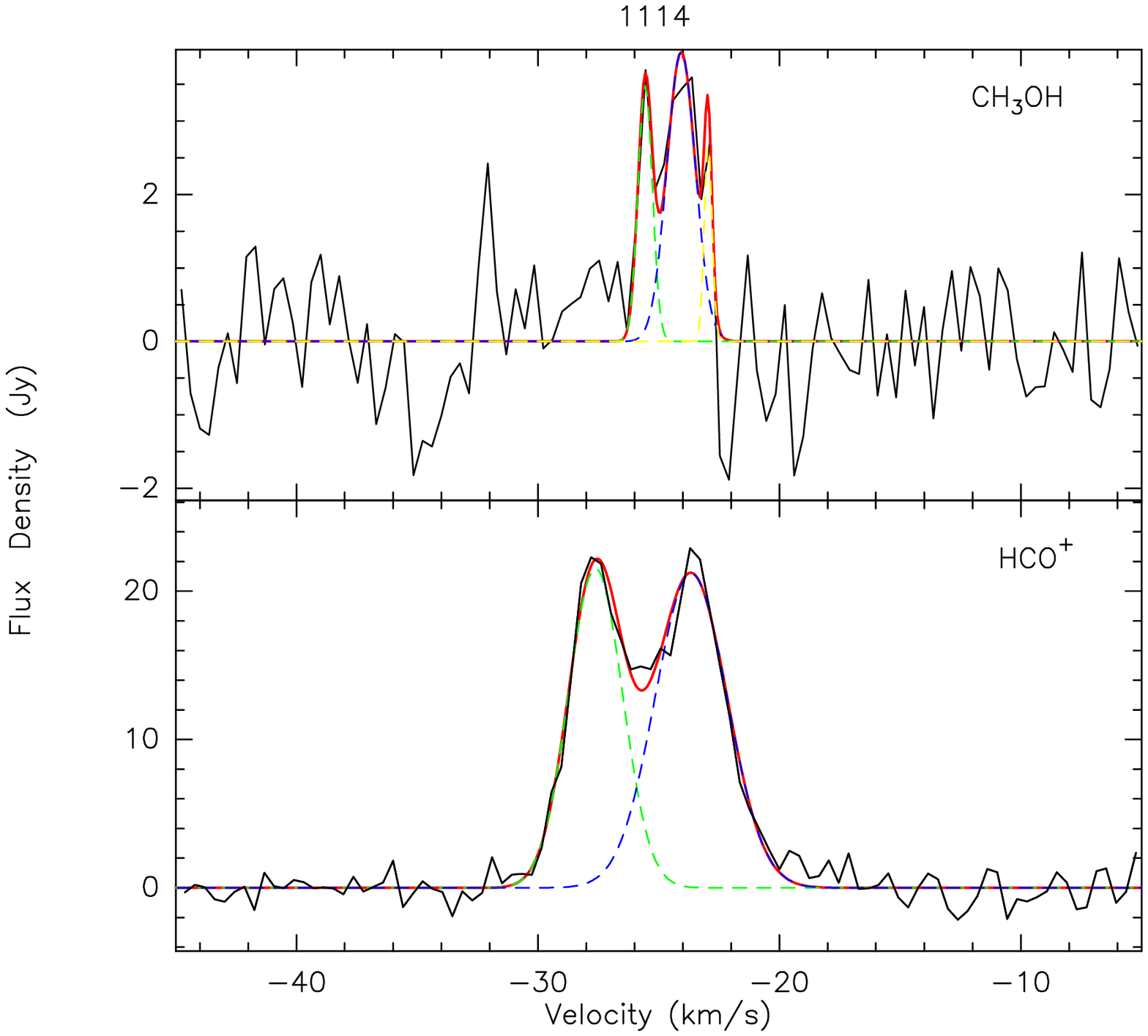}
\end{minipage}
\begin{minipage}[b]{4.1cm}
\includegraphics[scale=0.23,angle=-90]{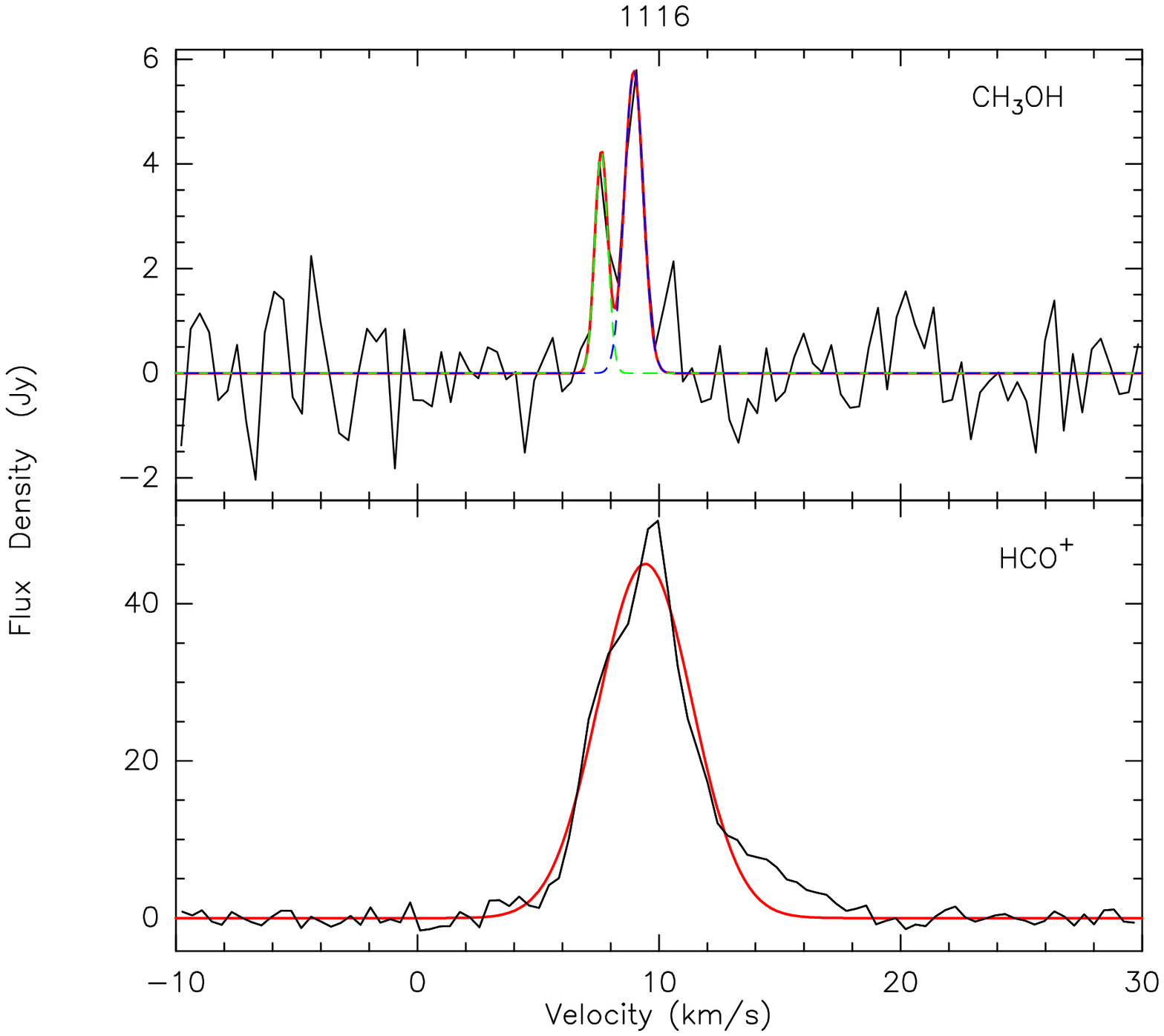}
\end{minipage}
}
\mbox{
\begin{minipage}[b]{4.1cm}
\includegraphics[scale=0.23,angle=-90]{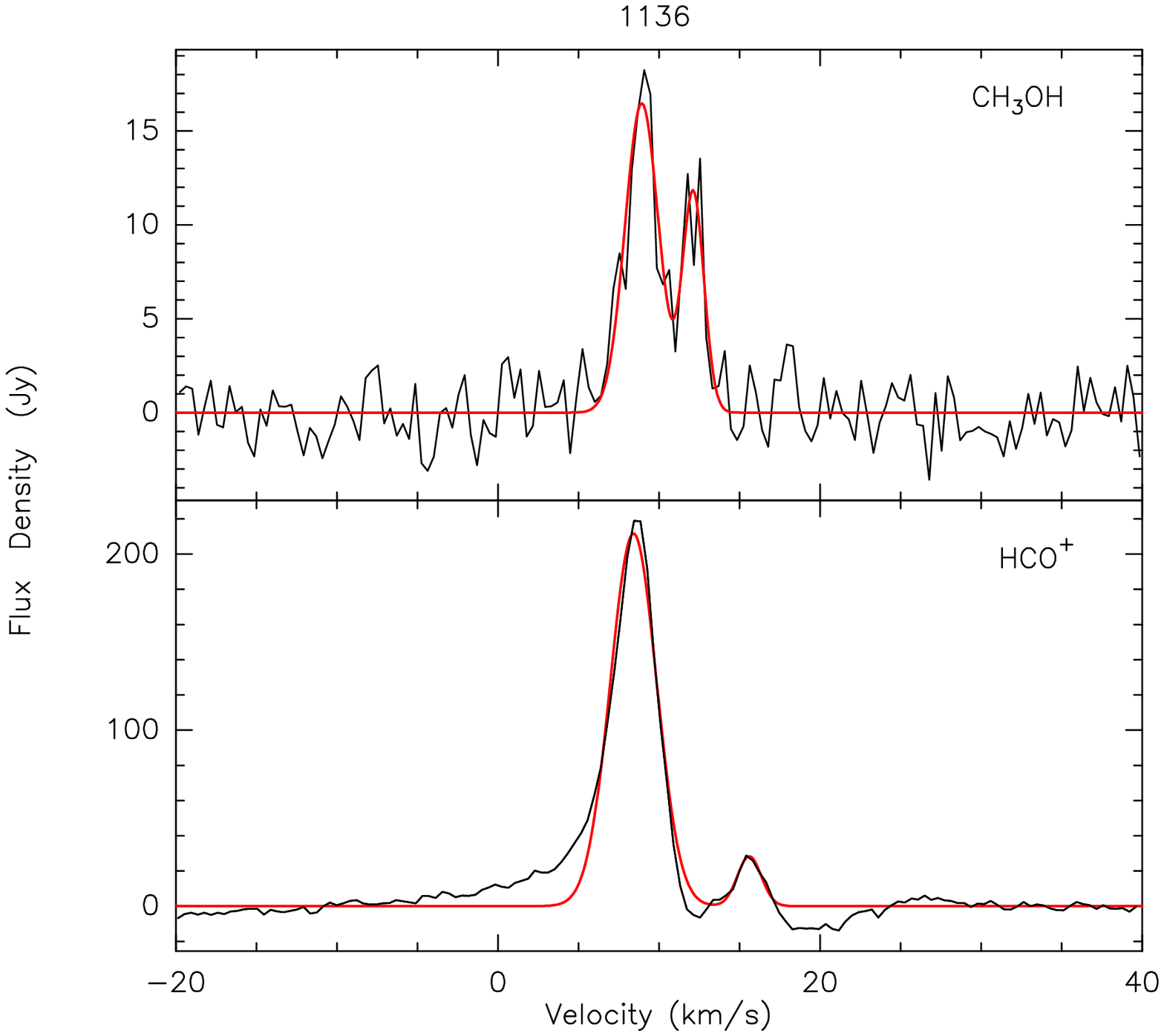}
\end{minipage}
\begin{minipage}[b]{4.1cm}
\includegraphics[scale=0.23,angle=-90]{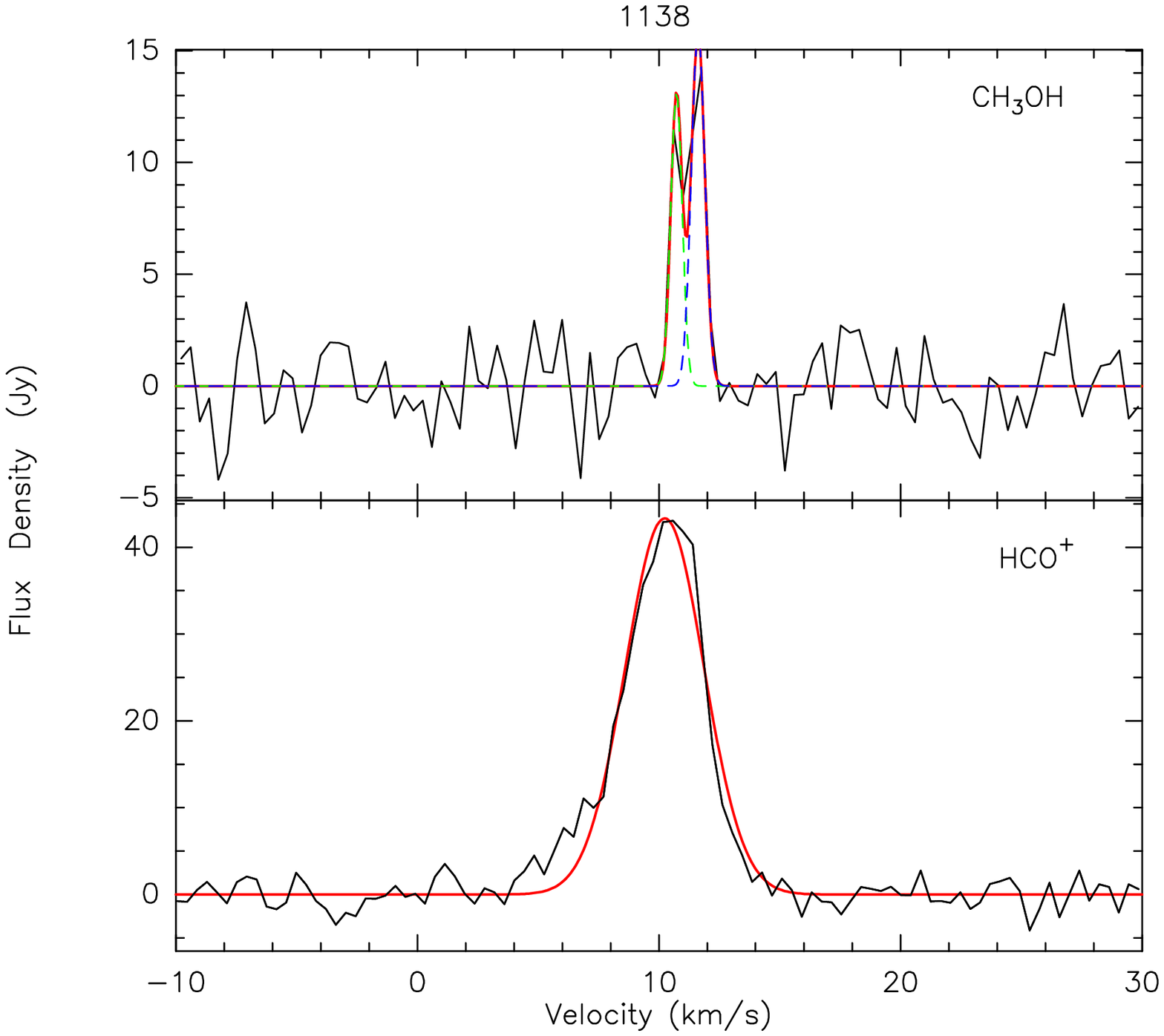}
\end{minipage}
\begin{minipage}[b]{4.1cm}
\includegraphics[scale=0.23,angle=-90]{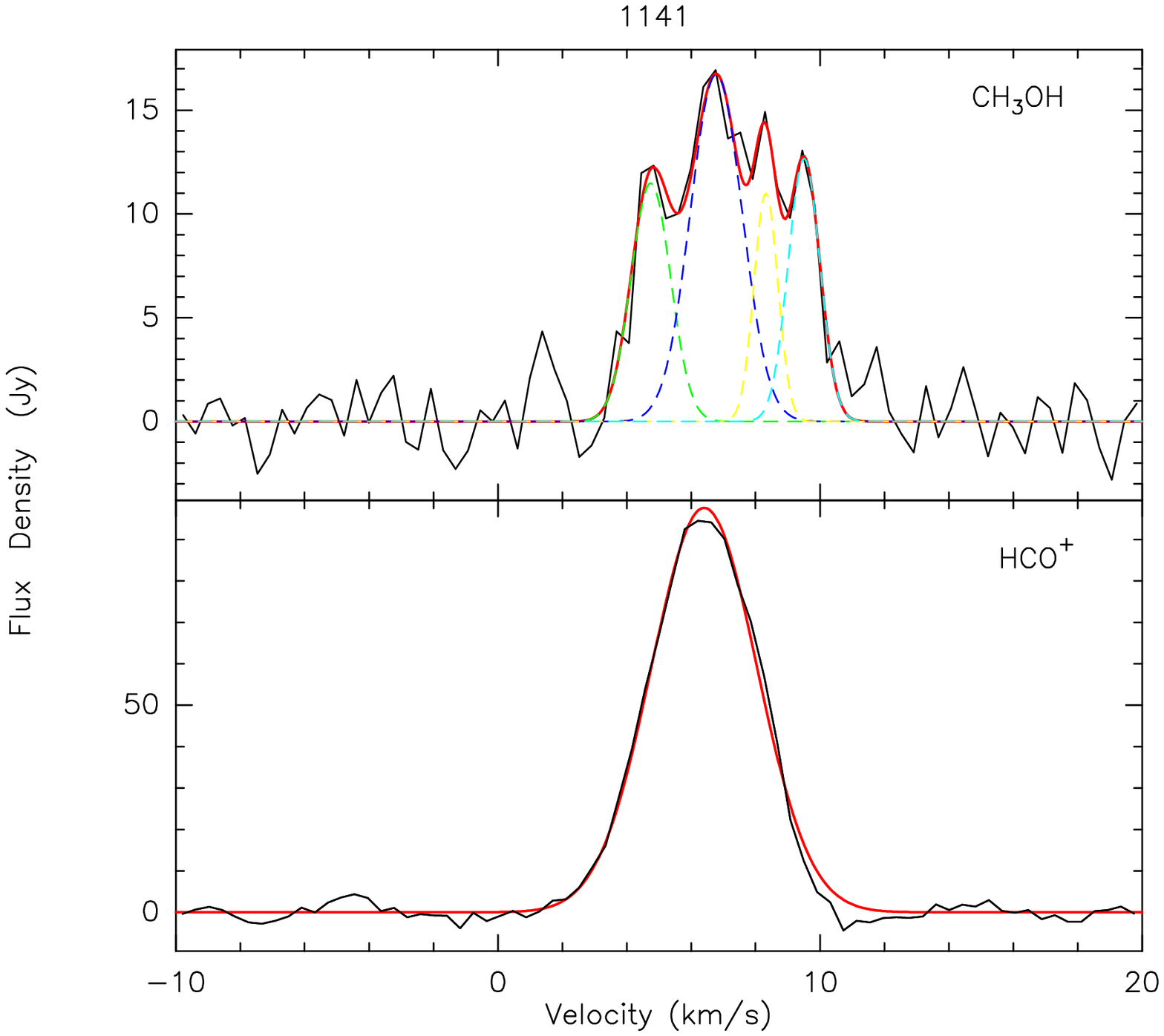}
\end{minipage}
\begin{minipage}[b]{4.1cm}
\includegraphics[scale=0.23,angle=-90]{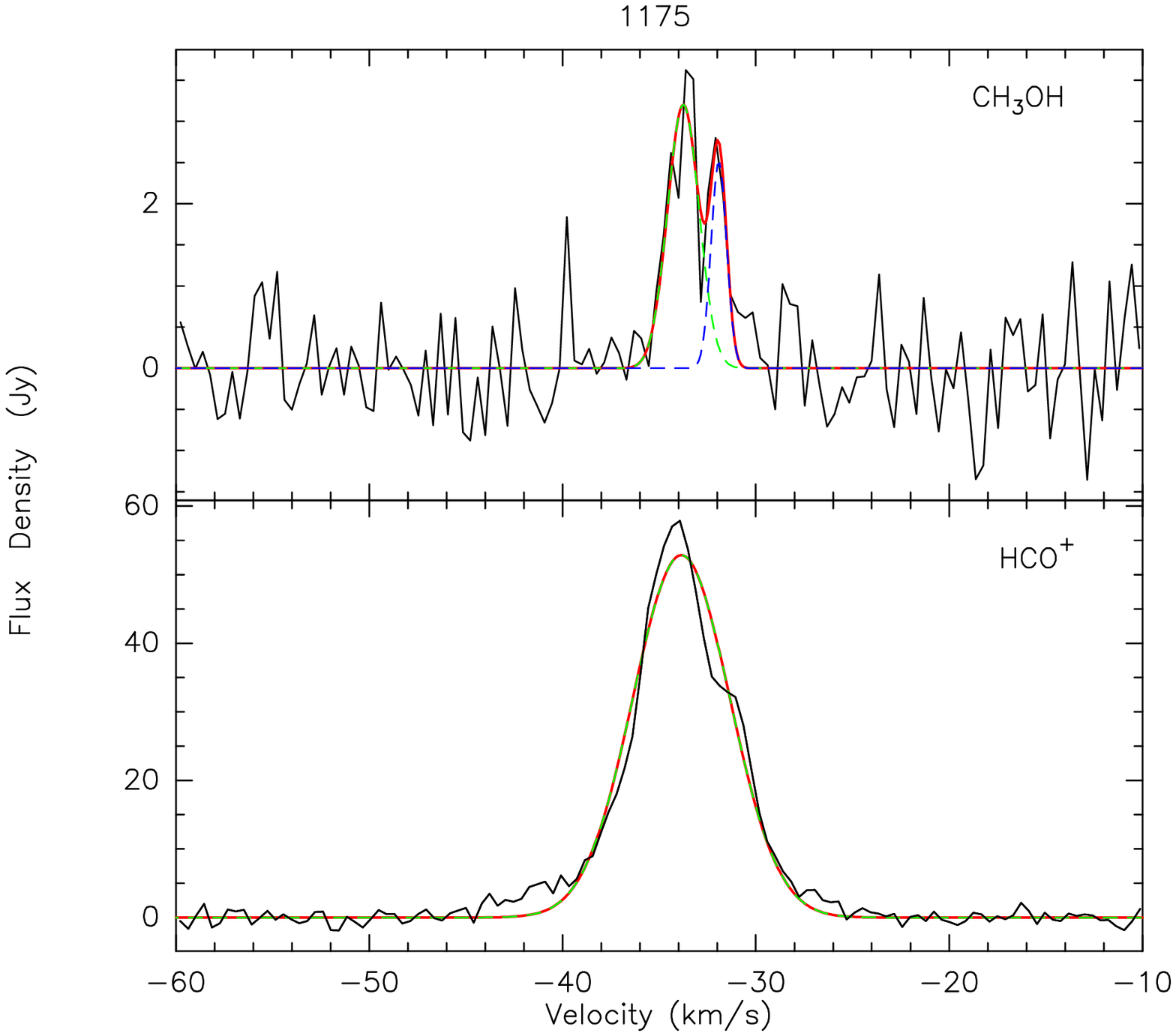}
\end{minipage}
}
\caption{The upper panel and lower panel show the spectra of the 95 GHz methanol and HCO$^+$ emission detected towards BGPS sources, respectively. The red bold line shows the sum of the Gaussian fitting results, while the colored dashed lines show the individual Gaussian fit components.\label{fig:masers}\\
(The complete figure set (213 images) is available in the online material.)}
\end{figure}

In total, 1020 BGPS sources were searched for 95 GHz class I methanol maser emission and detections were made towards 213 sources.  The spectra of the 95 GHz methanol and HCO$^+$ emission of the 213 detections are shown in Figure \ref{fig:masers}.
For the 95 GHz methanol emission, the peak flux densities of the detected emission derived from the Gaussian fits range from $\sim$ 0.7 to 269 Jy (corresponding to main beam temperatures $T_{\rm mb}$ $\sim$ from 0.02 to 11.4 K). The integrated flux densities obtained from all features for each source range from 2.1 to 460.6 Jy km s$^{-1}$. 
Some sources contain both strong narrow spectral features and weak broad emission which is consistent with the results of previous class~I methanol maser observations. For example, BGPS1810 was previously detected in \cite{2011ApJS..196....9C}, BGPS2601 in \cite{2000MNRAS.317..315V} and BGPS3594 in \cite{2005MNRAS.359.1498E}.  
The line width of individual 95 GHz emission components obtained from Gaussian fitting ranges from 0.39 to 12.08 km s$^{-1}$ with a mean of 1.62 km s$^{-1}$ and a median of 1.28 km s$^{-1}$. The spectra of the 95 GHz methanol emission in most sources contains one or more narrow spectral features, with approximately 49\% (104/213) of the sources having one or more components with line width $<$ 1 km s$^{-1}$ , approximately 85\% (181/213) of the sources having one or more components with line width $<$ 2 km s$^{-1}$, and approximately 98\% (208/213) of the sources contain one or more components narrower than 3 km s$^{-1}$.

Single-dish observations cannot reliably distinguish between maser and thermal emission in spectra with broad line widths. Interferometric observations \citep[e.g.,][]{2009ApJ...702.1615C} have been made towards sources with broad single dish spectral profiles in class I methanol transitions and detected some emission on arcsecond scales (consistent with it being maser emission).  For example, interferometric observations of the 95 GHz methanol transition towards W 51 by \cite{1993LNP...412..211P} shows it to be maser emission, while our single dish observations of BGPS6363 (the W 51 region) shows a broad line width of 8.08 km s$^{-1}$. 
The current single dish observations cannot distinguish whether sources which show only broad components are due to spectral blending of several weak masers or due to quasi-thermal emission. 
Some of these sources are clearly not single component, but are better fitted by a single broad component than by multiple narrow components (e.g. BGPS source 1250, 4539 and 4560).

To obtain a more reliable sample of 95 GHz methanol maser detections, we have undertaken a source by source comparison between the 95 GHz methanol emission and the HCO$^+$ emission (the strongest component at the systemic velocity) which was observed simultaneously.  Sources where the 95 GHz methanol emission has a line width narrower than that of the HCO$^+$thermal emission are likely to be masers.  Sources which do not satisfy this criterion are marked with a $B$ in Table \ref{tab:PCat} and are considered as maser candidates, requiring higher resolution observations for confirmation.  Using this criterion 205 of the 213 95 GHz methanol detections are considered masers, with the remaining 8 sources identified as maser candidates (marked with a * in the ref column in Tables \ref{tab:MCat} and  \ref{tab:PCat}).

It is worth noting that for five of the BGPS sources detected in the current observations we have fitted a single Gaussian profile, whereas previously they have been reported to contain multiple narrow spectral components.
BGPS sources 1314 and 1710 were previously detected with the Mopra telescope \citep{2013ApJS..206....9C,2011ApJS..196....9C} and the smaller beam size and southerly declination of these two sources mean that the previous observations are higher signal-to-noise than those reported here. Similarly, the observations of BGPS sources 1421, 6657 and 6955 by \cite{2013ApJ...763....2G}, have better signal-to-noise than our observations and so for the catalog we take the parameters from the published literature.

For the 807 BGPS sources that we observed but were found not to have 95 GHz methanol emission with a signal-to-noise ratio greater than 3, the observed positions and measured rms noise level are given in Table \ref{tab:807}.

\subsection{95 GHz Methanol Masers Detection Rate} \label{subsec:rate}
We detected a total of 205 sources with 95 GHz methanol maser emission stronger than 3$\sigma$ (the typical peak flux density of the detections is greater than 4 Jy). In total 1020 BGPS sources were searched corresponding to a detection rate of 20\%. Among the 205 detected 95 GHz methanol maser sources, 61 have been previously reported in the literature, while the remaining 144 are new detections. 

\begin{figure}
\center
\includegraphics[scale=0.4]{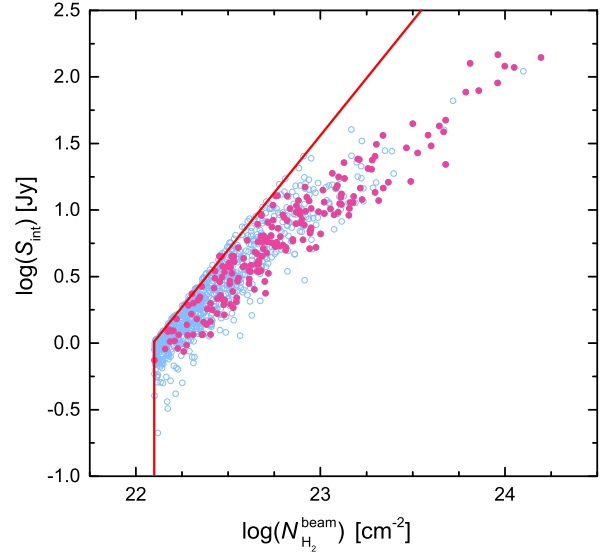}
\caption{The integrated flux densities vs. beam-averaged H$_2$ column density plotted on a log-log graph for 982 BGPS sources with and without 95 GHz methanol masers detections (marked by magenta solid circles and blue hollow circles, respectively). The red lines mark the criterion given by \cite{2012ApJS..200....5C}.\label{fig:criteria}} 
\end{figure}

We detected 204 95 GHz methanol masers towards the 982 BGPS sources selected on the basis of criteria proposed by \cite{2012ApJS..200....5C}, which use the beam-averaged column density and integrated flux density from the BGPS catalog. This corresponds to a detection rate of 21\% for this sample.  
The detection rates of previous 95 GHz methanol masers surveys are summarized in Table \ref{tab:sum}. The 21\% overall detection rate of this survey demonstrates that the BGPS-based selection criteria we have used is a good method for finding class I methanol masers, but not as efficient as EGO-based search criteria \cite[i.e. 55\% and 70\% in][respectively]{2011ApJS..196....9C,2013ApJS..206....9C}. EGOs are well established as sources which contain massive young stellar objects with ongoing outflows, so a close association with class~I methanol masers is not surprising, however, the detection rate achieved towards the current BGPS sample is significantly lower than predicted by \cite{2012ApJS..200....5C}.  
The likely reason for this is that the sources used to establish the original criteria were associated with both a BGPS source and a GLIMPSE point source meeting specific criteria.
The GLIMPSE point sources were selected to have mid-infrared colors similar to those of sources known to be associated with class I and class II methanol masers from \cite{2006ApJ...638..241E}.
However, in our sample there are many BGPS sources without an associated GLIMPSE point source.
Moreover, the limited sensitivity and large beam size for the current observations may lead to non-detection of weak masers.  
In a log-log graph of BGPS flux density versus BGPS beam-averaged column density, the sources in \cite{2012ApJS..200....5C} with and without the 95 GHz methanol maser detections can be distinguished. However, Fig.~\ref{fig:criteria} shows that in our sample, a significant fraction of the BGPS sources that satisfy these criteria do not have an associated maser.
Despite this, the current survey has detected 147 new class~I methanol maser sources, which have not been detected in previous EGO-based surveys, or those using other selection criteria.  Extending the sample of known class~I methanol maser sources is important to better understand the range of environments which are capable of producing class I methanol masers.

Five sources (BGPS source 1412, 2096, 2292, 3208 and 5373) were previously observed by \cite{2012ApJS..200....5C}, thus for a total of 1015 sources we can investigate the statistical classification techniques of \citet{2016PASA...33...15M}. \citeauthor{2016PASA...33...15M} identified 490 of these 1015 sources as having a probability of 0.5 or more of being associated with a 95 GHz methanol maser by at least one of the statistical classification techniques.  Of these 490 BGPS sources 155 have 95 GHz methanol maser emission (these sources are indicated with a superscript $M$ in Table \ref{tab:MCat}), corresponding to a detection rate of 32\%. The 7 maser candidates and the other 328 undetected sources are indicated with a superscript $M$ in Table \ref{tab:PCat} and Table \ref{tab:807}.

For comparison with the predictions of \cite{2016PASA...33...15M}, we have summarised the results for each of the three classification techniques using a cross-validation table (Table \ref{tab:classification_results}). 
If a classification technique successfully identify a source with or without associated masers, the source is counted as a ``true positive" or a ``true negative". A ``false positive" occurs when a source without a maser association is predicted to be associated with a maser, while a ``false negative" outline is when a source with a maser association is not predicted to be associated by the classification technique.
The sensitivity, or true positive rate, is the percentage of maser associations correctly predicted by the model, and the specificity, or the true negative rate, is the percentage of maser non-associations correctly predicted. The accuracy is the proportion of true results in a population and it measure the veracity of any diagnostic test.

Regarding the utilization of these classification techniques, our results are in agreement with \cite{2016PASA...33...15M} in several aspects.
Firstly, in a larger sample random forests showed the highest sensitivity, while LDA had the lowest but with the highest specificity. And in this case, the accuracy of random forests was not superior to that of parametric methods.
This supports the idea that random forests has the capability to outperform parametric methods for larger datasets, but that perhaps more training data is required.
Also, our results show that LDA has higher specificity than normalised LDA (i.e. LDA using the ``nomalised" data transformed by a log function), while normalized LDA achieved higher sensitivity. But in our case, using the transformed data slightly decreased the accuracy of LDA. One possible reason is that some of the non-detected sources have weak maser emission, lower than the sensitivity limit of the current observations (and indeed, normalised LDA had more ``false positives").

In Table \ref{tab:classification_results}, the 7 maser candidates are counted as ``false positives". 
It is worth noting that 6 out of the 7 maser candidates are predicted to be masers with great probability (very close to 1.00) by all statistical methods, except BGPS2612 which is predicted by three of the methods.  This suggests that perhaps maser and thermal emission are blended in these extreme environments to produce the broad line widths detected by our single-dish observations.

\begin{figure}
\center
\plottwo{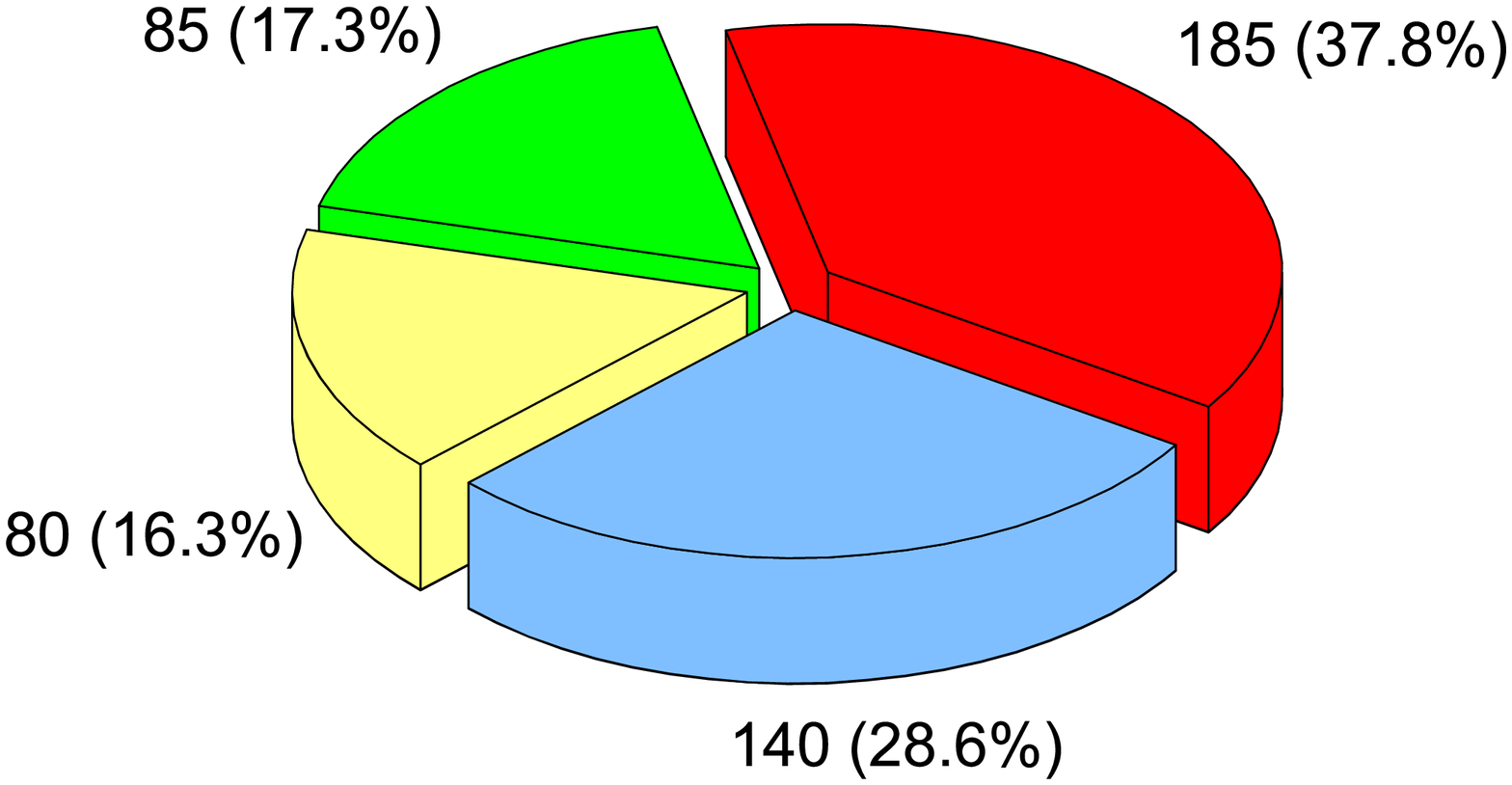}{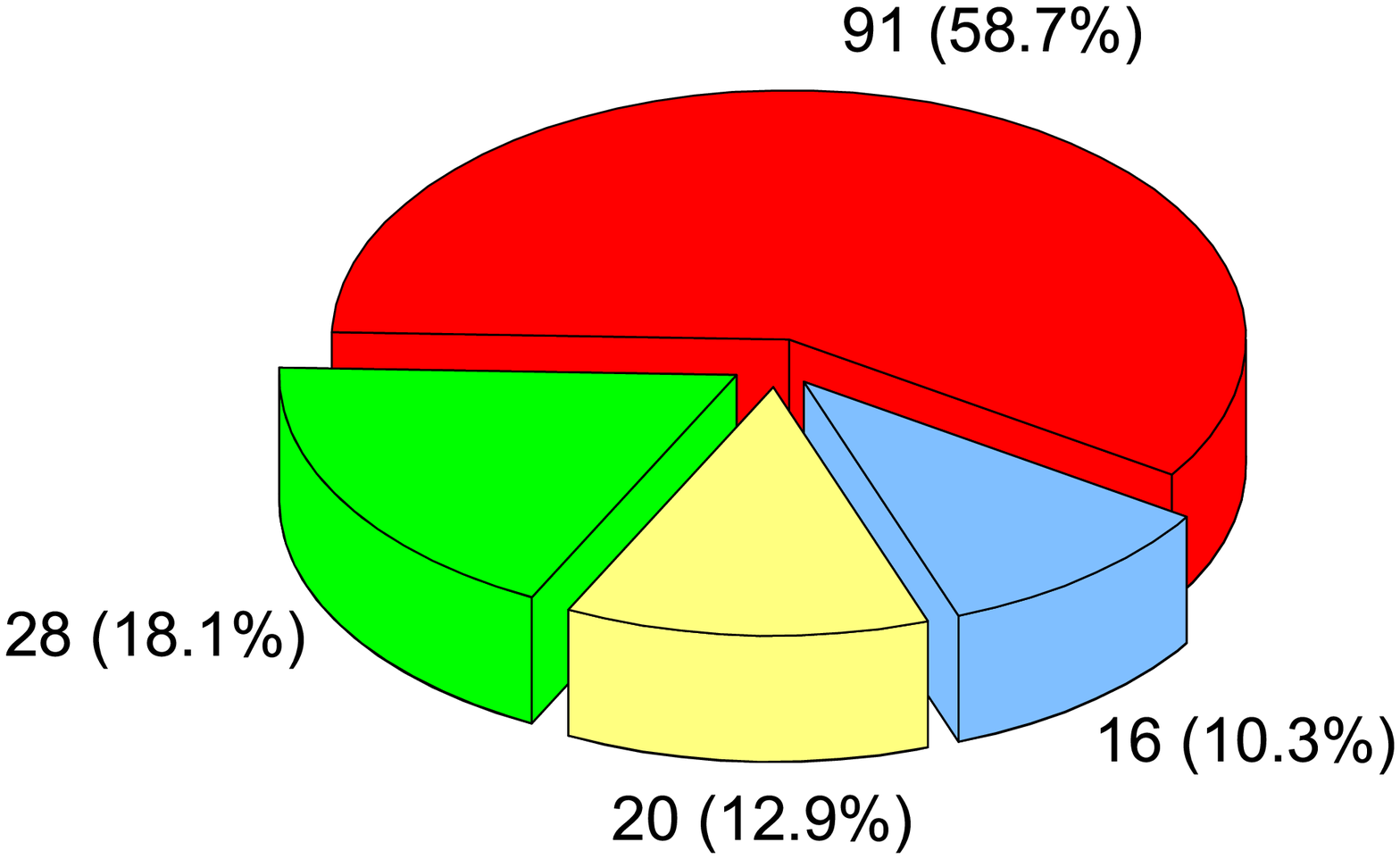}
\caption{The percentage of 490 BGPS sources (left) and 155 maser-detected sources (right) that predicted by different numbers of statistical methods. The blue, yellow, green and red parts mean BGPS sources are predicted by one, two, three or all kinds of methods, respectively. \label{fig:pie}}
\end{figure}

It is worth noting that the majority of the 739 BGPS sources which were listed by \citet{2016PASA...33...15M} as likely to have an associated 95~GHz methanol maser were predicted to have a maser by more than one of the statistical classification techniques.  The left-hand panel of Fig.~\ref{fig:pie} shows the percentage of the 490 BGPS sources from the \citeauthor{2016PASA...33...15M} we observed that were predicted by one (blue), two (yellow), three (green) or all four (red) classification methods.  The right-hand panel of the same figure shows the percentages for the 155 BGPS sources with a maser-detection.  Combining the information in these charts we see that the detection rate for BGPS sources which were only predicted to be a maser by one model is 16 of 140 ($\sim$11\%), while for those predicted by all four classification techniques it is 50\%. 
 
If we wanted to increase the efficiency of any search, then by restricting our sample to only include BGPS sources predicted to be a maser by either 3 or 4 of the different classification techniques, they would detect $\sim$77\% of the total number of 95 GHz maser sources. These sources will be the priority targets for future class I methanol maser searches.

\subsection{HCO$^+$ Emission} \label{subsec:HCO}

In general, for maser emission associated with star formation regions, we expect to detect thermal emission from a variety of species originating from dense molecular clumps in the vicinity \citep{1993ApJ...417..645G}. After consideration of the beam size of the PMO 13.7 m radio telescope and the strength of the molecular lines at frequencies close to the 95 GHz methanol transition we chose to observe the HCO$^+$ ($J$=1--0) transition to determine the velocity range of the thermal gas from the star forming region.

Parameters of Gaussian fits to HCO$^+$ emission for BGPS sources where we detected 95 GHz methanol emission are given in Table \ref{tab:HCO}. 
Given the angular resolution of the observations we expect to sometimes have contamination from foreground/background emission along the same line of sight as the target BGPS source.  However, for many sources (e.g. BGPS1465, 2152) the HCO$^+$ emission is observed to not show a simple Gaussian profile, exhibiting multiple components peaks or asymmetries, which result in poor Gaussian fitting results and suggest the presence of multiple dense gas clumps in these regions. Furthermore, many sources (e.g., BGPS5853, 6872) show broad wings which, in many cases, are indicative of outflows. The spectra from single pointing, single dish observations cannot distinguish whether the wings are due to outflow, or other phenomena, such as turbulence, rotation, collapse or shock. Determining the nature of the broad wings observed in some of the HCO$^+$ sources will require future mapping observations.

Many of the BGPS sources which do not have a detected 95 GHz methanol maser do show a clear detection in thermal HCO$^+$ emission, enabling us to make comparisons between the HCO$^+$ emission in BGPS sources with and without an associated methanol maser.  The sample of BGPS sources without associated 95 GHz methanol masers consists of the 625 sources BGPS sources without a methanol maser (from 807 maser non-detection sources in total) that contain no HCO$^+$ absorption and for which we have good Gaussian fitting of the HCO$^+$ spectrum (the parameters are listed in Table \ref{tab:807HCO}). As the primary purpose of the survey was to search for methanol masers, we did not make additional observations to improve the quality of the HCO$^+$ data in those sources without a maser detection.  The large number of maser non-detection BGPS sources for which we have good quality HCO$^+$ data (86\% of the sample) means that the statistical comparison is expected to be robust, as there is no obvious reason as to how, or why, the incompleteness would introduce bias.

\begin{figure}
\center
\includegraphics[scale=0.24]{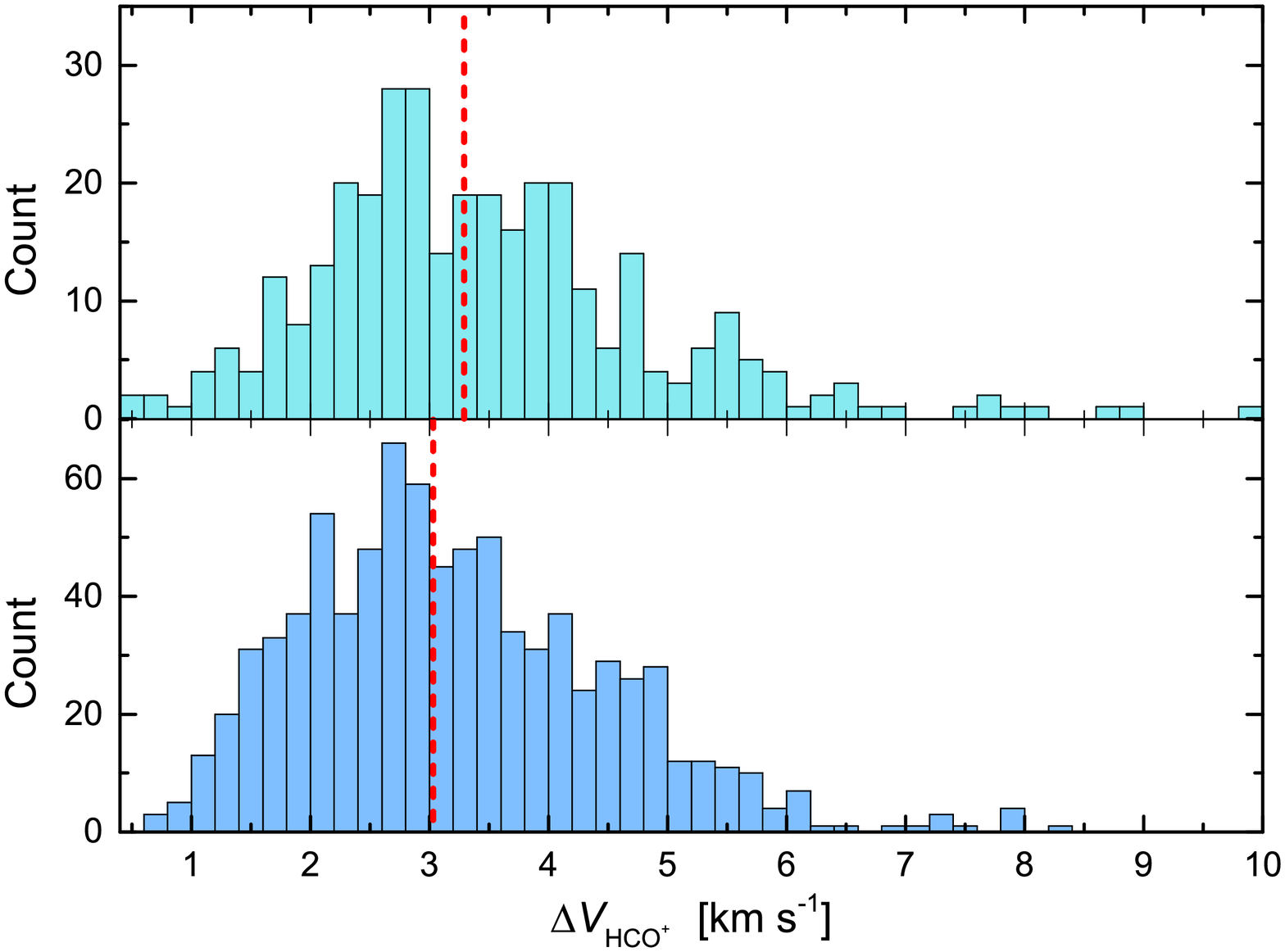}
\includegraphics[scale=0.24]{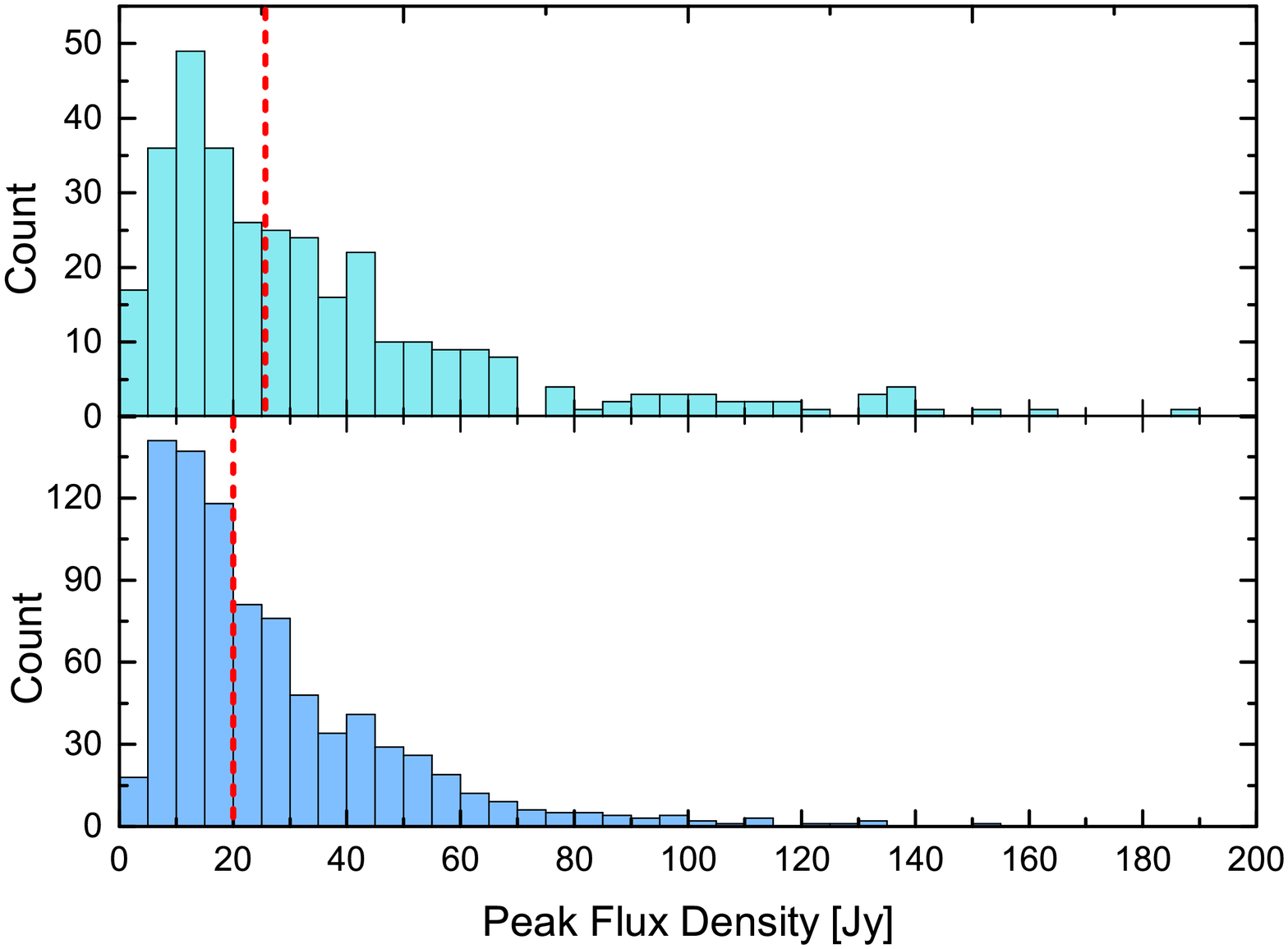}
\includegraphics[scale=0.24]{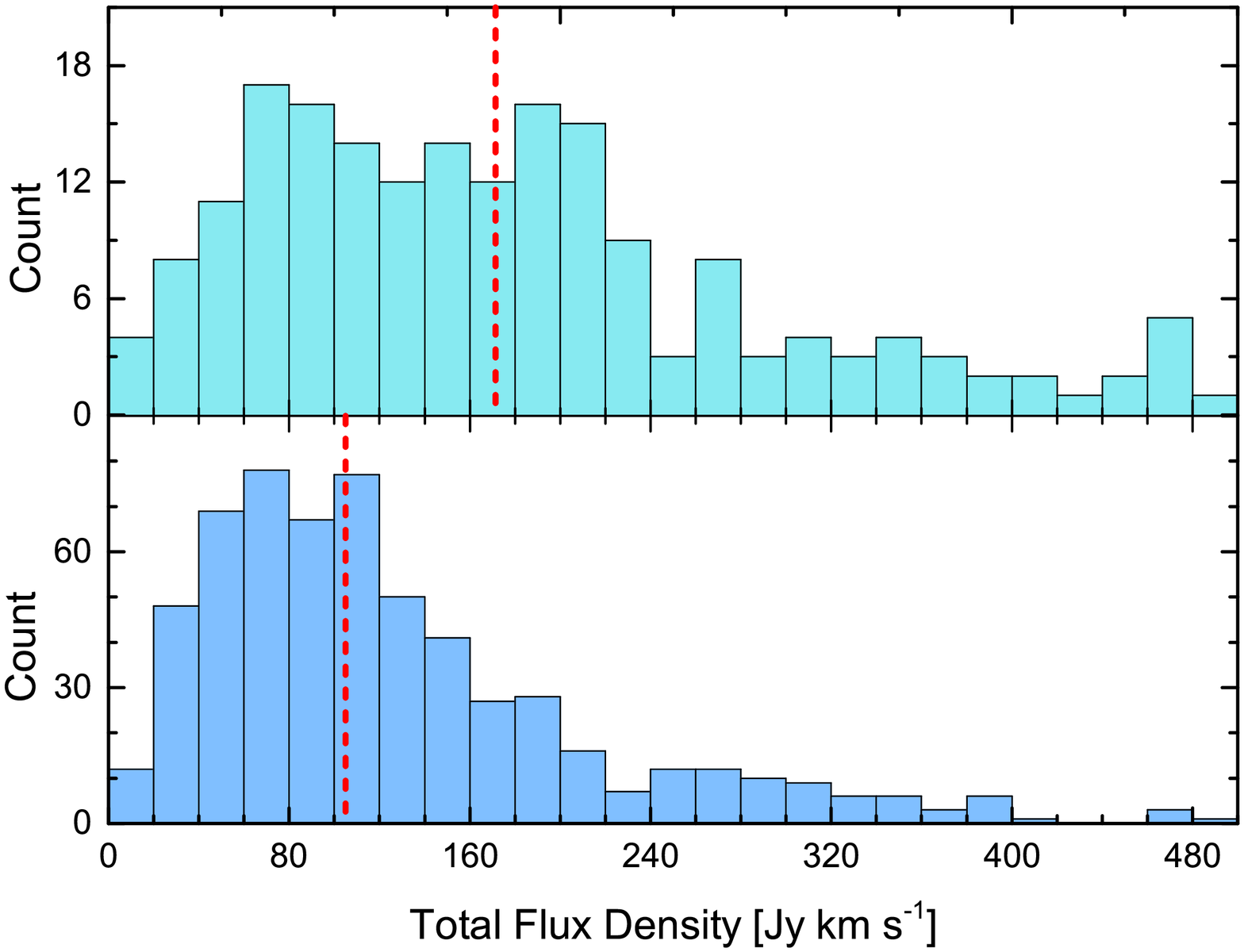}
\caption{The distribution of HCO$^+$ emission line parameters, width (left), peak flux density (middle) and total integrated flux density (right). The upper panels show the distribution for BGPS sources with an associated 95 GHz methanol masers and the lower panels show the distribution for those without. The median of each distribution is marked by a vertical red dashed line. \label{fig:hco_3}} 
\end{figure}

Figure \ref{fig:hco_3} shows a comparison of distribution of the line width, peak flux density and total integrated flux density for the HCO$^+$ emission between BGPS sources with and without associated 95 GHz methanol masers. The values are also listed in Table \ref{tab:HCO-compare}. A Kolmogorov-Smirnov test shows that the difference between the two distributions is statistically significant in all cases. For each of these quantities the BGPS sources with an associated class I methanol maser shows higher values in the HCO$^+$ emission, suggesting that regions which are hotter and where the star formation activity is stronger are more likely to generate masers.

Several BGPS sources with an associated methanol maser show absorption in the HCO$^+$ spectrum (e.g. BGPS6363 and BGPS6901 in Fig.~\ref{fig:masers}).  For these sources we made reference observations at several different pointings, but this did not change the absorption, demonstrating that it is not due to emission in the reference location. We did not detect HCO$^+$ emission in the velocity range from --200 km s$^{-1}$ to 200 km s$^{-1}$ towards the source BGPS6202, but both the current observations and \cite{2011ApJS..196....9C} detected 95 GHz methanol maser emission towards this source, so we consider it as a maser in our discussion below. It is worth noting that the absence of HCO$^+$ emission suggests that this is an atypical class~I methanol maser source and an interesting target for further investigation.

\subsection{the \textit{V}$_{\rm lsr}$ Difference between Methanol Masers and HCO$^+$} \label{subsec:Vlsr}

\begin{figure}
\plottwo{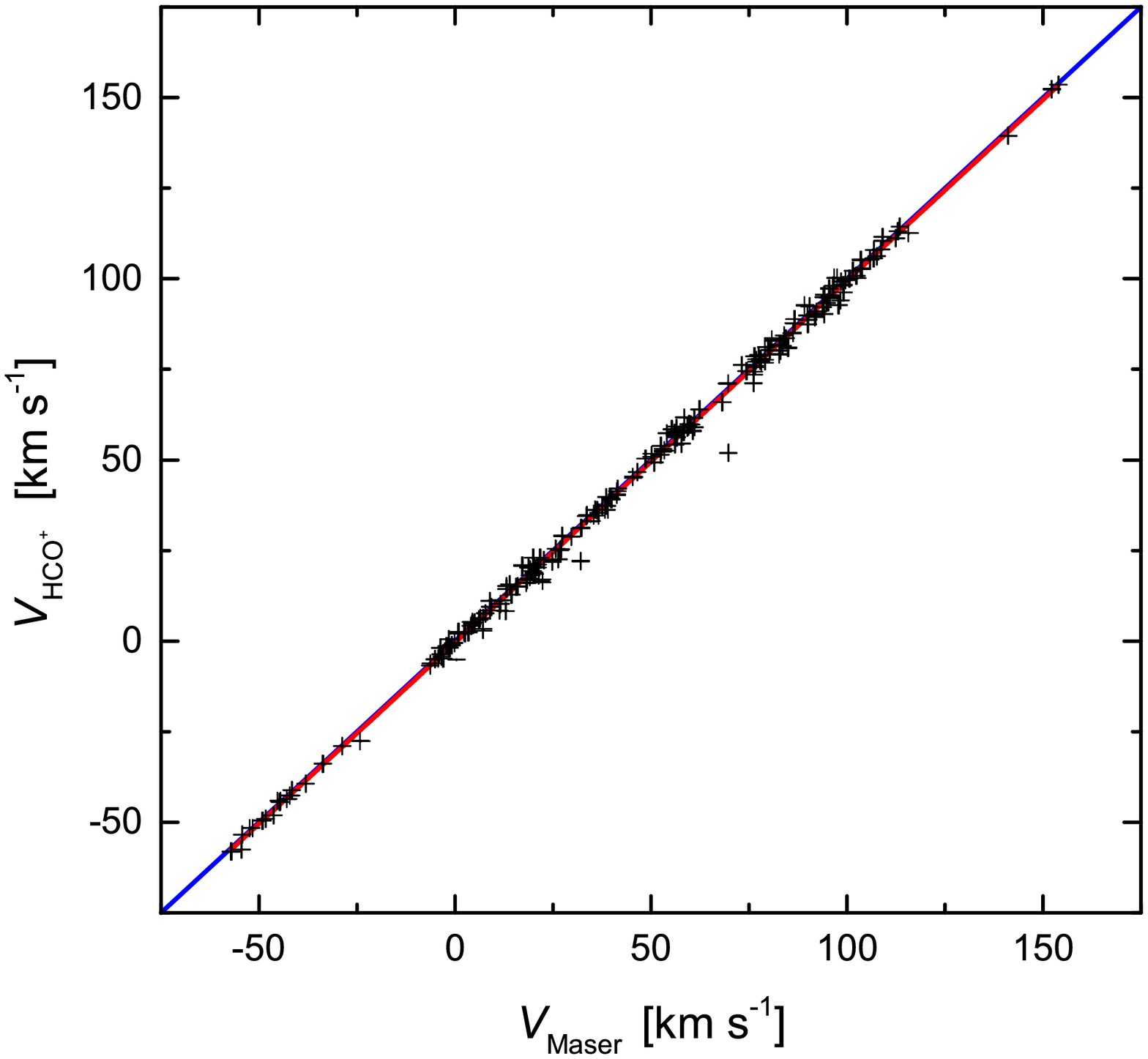}{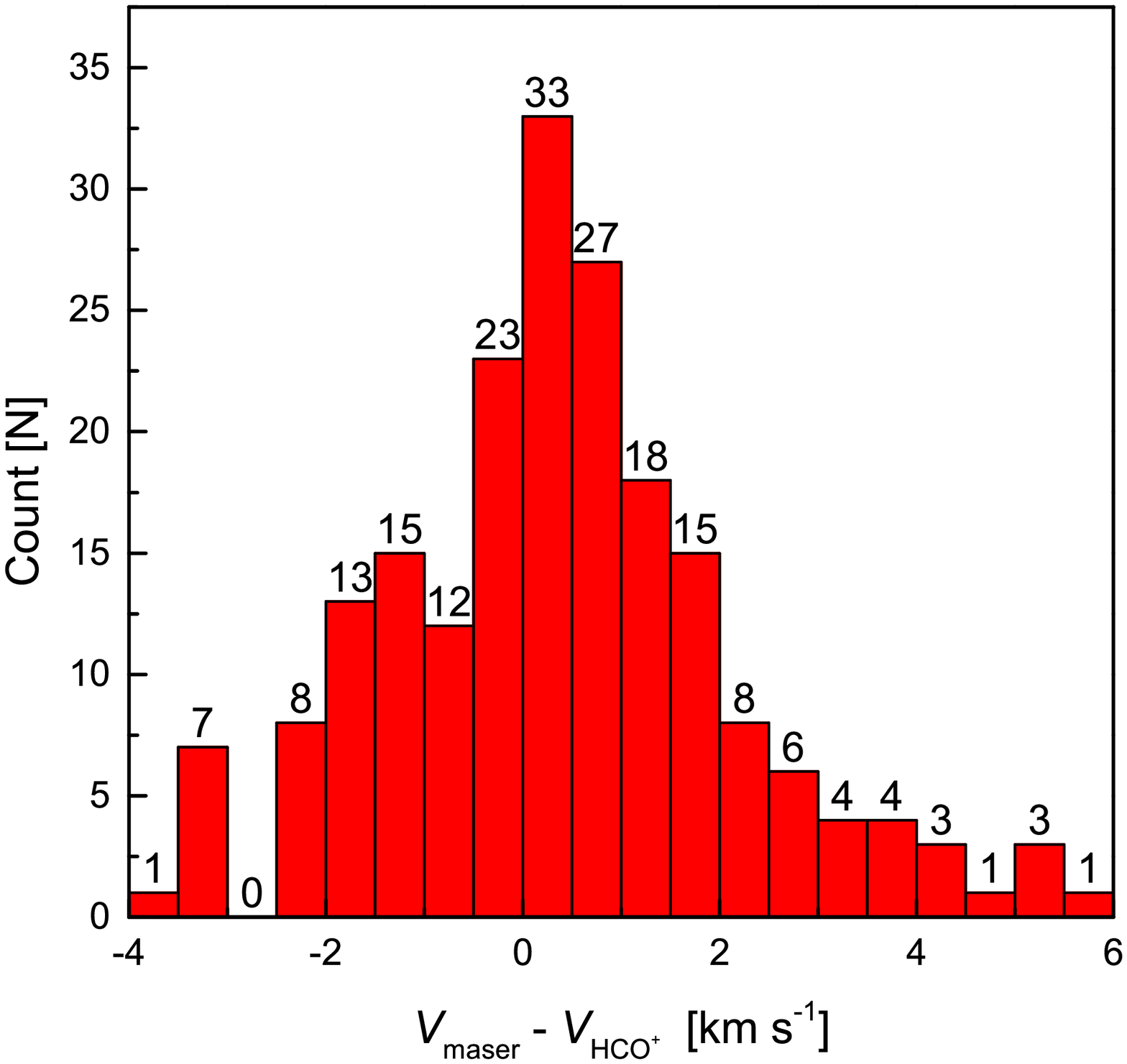}
\caption{(\textit{Left} panel) The $V_{\rm lsr}$ of the peak of the HCO$^+$ emission versus the peak of the 95 GHz methanol maser emission. The red line shows the best fit from linear regression analysis while the blue line is a line through the origin with a slope of 1. (\textit{Right} panel) A histogram (0.5 km s$^{-1}$ bins) showing the $V_{\rm lsr}$ difference between the 95 GHz methanol maser and HCO$^+$ emission peaks.  The two outlying sources (BGPS 2111 and 2386) are not included in the histogram.} \label{fig:lsr}
\end{figure}

In general, the observed velocities of class I methanol masers are no more than a few km s$^{-1}$ from the ambient molecular clouds \citep[e.g.,][]{1990A&A...240..116B, 2006MNRAS.373..411V}. The left-hand panel of Fig.~\ref{fig:lsr} shows the peak velocity $V_{\rm lsr}$ of the 95 GHz methanol masers versus that of the HCO$^+$ emission for 204 BGPS sources (i.e. all maser detections except for BGPS6202 which does not have detected HCO$^+$ emission). The red line shows the best fit from linear regression analysis, which has a slope of 0.999, standard error of 0.003 and a correlation coefficient of 0.998.  Thus the data is consistent with the blue line (Fig.~\ref{fig:lsr} left panel), which has a slope of 1 and passes through the origin (representing no difference between the $V_{\rm lsr}$ of the 95 GHz methanol maser and HCO$^+$ peaks).  There are two sources which can be see to deviate significantly from the line of best-fit, BGPS2111 (G14.78--0.33) and BGPS2386 (G18.26--0.25) which have velocity differences between the peak maser and HCO$^+$ emission of 10 km s$^{-1}$ and 17.8 km s$^{-1}$, respectively.  From Figure~\ref{fig:masers} we can can see that for BGPS2111 the HCO$^+$ emission shows two components, with strong emission at 22 km s$^{-1}$, and a weaker peak at 30 km s$^{-1}$, which correspond to near kinematic distances of 2.2 kpc and 2.8 kpc, respectively.  The peak of the maser emission is at 32~km s$^{-1}$ and it is likely that the maser is associated with this weaker HCO$^+$ component. Using  a Bayesian Distance Calculator \footnote{\url{http://bessel.vlbi-astrometry.org/bayesian}}, the 22 km s$^{-1}$ component of HCO$^+$ is most likely located in the Sagittarius Arm (probability 0.55), while the 30 km s$^{-1}$ component of HCO$^+$ (and maser) are likely in the Scutum Arm (probability 0.8). Likewise, for BGPS2386, the maser and one of the HCO$^+$ components (at about 71~km s$^{-1}$) are likely within the Norma Arm (probability 0.7), although the strongest HCO$^+$ emission is at a significantly lower velocity of 52~km s$^{-1}$. We cannot rule out the possibility that all the HCO$^+$ and methanol emission arises from a single site with a systematic velocity equal to the strongest HCO$^+$ emission.  However, given the large number of sources in our sample and the intrinsically non-linear nature of maser emission it is not surprising that some sources have multiple HCO$^+$ sources along the line of sight and that in some cases the masers are associated with an HCO$^+$ component of lower flux density.  
There are also counter examples, such as BGPS6712 and 7502, where there are multiple maser features but only one HCO$^+$ component, this may result from multiple outflows or shocked regions within a single star formation region. 

It should be noted that in BGPS6712 and BGPS7502 there are maser components offset from the HCO$^+$ peak velocity by about 15 km s$^{-1}$ and 4 km s$^{-1}$, respectively. A similar phenomenon has been reported by \cite{2010MNRAS.408..133V} who have identified a 44 GHz class~I methanol maser with a component offset from the systemic velocity by 30 km s$^{-1}$.  
\citeauthor{2010MNRAS.408..133V} suggest that this is due to an outflow parallel to the line of sight. Despite neither of the sources identified here having as large a velocity offset, BGPS6712 may represent the second most extreme case of this sort of phenomenon. 

The right-hand panel of Fig.~\ref{fig:lsr} shows the distribution of the difference between the maser and HCO$^+$ peak velocity for 202 sources (all except for BGPS2111, 2386 and 6202).  For nearly half of the sources (95/204$\sim$47\%), the velocity difference is less than 1 km s$^{-1}$. 
Class I methanol masers are associated with outflows in the majority of the cases, and as such it might be expected that they may peak at velocities offset from the systemic velocity of the source powering the outflow.  However, the very small velocity difference between the maser peak and thermal gas shows that the offset is generally small.

This panel also shows that there are more sources for which the maser emission is red-shifted with respect to the thermal emission than those which are blue-shifted (123/79 $\sim$ 60\%:40\%). This perhaps suggests a slight preference for the class I maser emission to be seen in outflows which are directed more away, rather than towards the observer, however, it is not clear if this is a statistically significant result.

\section{Searches for 95 GHz Methanol Masers: 1994--2016} \label{sec:catalog}

The 8$_{0}$--7$_{1}$A$^{+}$ 95 GHz class I methanol maser transition was first identified by \cite{1968JChPh..48.5299L}, with the first astronomical detections by \cite{1986Ap&SS.118..405O} towards Orion--KL and \cite{1986PASJ...38..531N} towards S 235. Subsequently, there were several interferometric observations, such as 
\cite{1988ApJ...330L..61P,1990ApJ...364..555P} which confirmed that the transition was exhibiting maser emission. However, there were no systematic searches for the 95 GHz maser transition until \cite{1994AAS..103..129K}. We have combined the results of all previous 95 GHz methanol masers single-dish searches which are reported in \cite{1994AAS..103..129K,1995AZh....72...22V,2000MNRAS.317..315V,2005MNRAS.359.1498E,2006ARep...50..289K,2010AA...517A..56F,2011ApJS..196....9C,2012ApJS..200....5C, 2013ApJ...763....2G,2013ApJS..206....9C} to produce a catalog of all maser sources (line width $<$ 3 km s$^{-1}$) in this transition. 

\begin{figure}
\center
\includegraphics[scale=0.4]{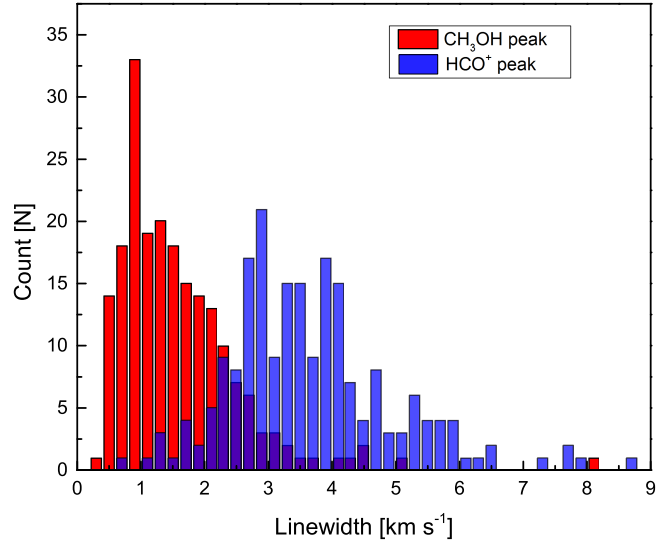}
\caption{The distribution of linewidth between peak maser emission and HCO$^+$ thermal emission in the current observations. \label{fig:linewidth}} 
\end{figure}

To ensure that the maser catalog does not include sources which have only thermal emission, we have undertaken a source by source comparison between the peak 95 GHz methanol emission and the HCO$^+$ emission (the strongest component at the systemic velocity), which was observed simultaneously. Figure \ref{fig:linewidth} shows the distribution of linewidths of all the 204 methanol masers (except for BGPS6202 which does not have detected HCO$^+$ emission), and HCO$^+$ emission. The median and average of the methanol masers linewidths are 1.37 km s$^{-1}$ and 1.58 km s$^{-1}$, respectively. Demonstrating that the maser emission linewidths are relatively narrow.  In contrast the median and average of the HCO$^+$ emission linewidths are  3.47 km s$^{-1}$and 3.65 km s$^{-1}$, respectively. 
The linewidths of the HCO$^+$ emission in most sources is in the range 2.5 - 4.5 km s$^{-1}$ and similar results are observed in thermal CS and HNC emission associated with class I methanol masers \cite{2013ApJS..206...22C}.  For linewidths greater than 2.5 km s$^{-1}$ there are fewer methanol sources than HCO$^+$ source.
Over 85\% of the methanol maser sources have linewidths narrower than 2.5 km s$^{-1}$ and \cite{2013ApJS..206....9C} also found 2.5 km s$^{-1}$ a typical upper limit for maser components.  So for the 95 GHz methanol detections from the literature for which we do not have information from thermal lines towards the same location we consider those with linewidths $<$ 2.5 km s$^{-1}$ to be masers (Table \ref{tab:MCat}) and those with linewidths $>$ 2.5 km s$^{-1}$ to be maser candidates.  The 37 identified maser candidates are listed separately in Table \ref{tab:PCat}.

Apart from some sources (see Section \ref{subsec:detection}), for the majority of the sources which have been observed by multiple authors, we use the line parameters from the most recent publication, as these are generally more reliable due to advances in instruments and telescope design. 
However, accurate positions have not been determined for the majority of 95 GHz methanol masers, so it is possible that in some cases the pointing of earlier observations may be closer to the real location than the most recent detection.\\

The catalog of 95 GHz methanol masers and the list of maser candidates contain the following information :\\
Column 1: Source name, arranged by increasing Right Ascension. Those 155 sources Table \ref{tab:MCat} and 7 sources in Table \ref{tab:PCat} given a superscript $M$ were predicted as having a greater than 50\% chance of being a maser by at least one of the statistical classification techniques utilized by \cite{2016PASA...33...15M}. \\
Column 2--3: The equatorial coordinates for each source. \\
Columns 4--7: The velocity at peak $V_{\rm lsr}$, the line width $\Delta V$, the peak flux density $P$, the integrated flux density $S$ and corresponding fitting error for each of the maser features that have been estimated from Gaussian fits. \\
Column 8: The total integrated flux density $S_{\rm int}$ of the maser spectrum obtained by adding the integrated flux density of all maser features in the source. \\
Column 9: 1$\sigma_{\rm rms}$ noise of the observations. \\
Column 10: The distance for each maser, mainly estimated from the Galactic model of \cite{2014ApJ...783..130R}, with the exception of some sources with parallax distance measurements \citep{2008PASJ...60..961H,2008PASJ...60...37H,2014ApJ...783..130R,2016arXiv161000242X} marked by $P$. Sources marked a superscript $G$, $D$, $S$ or $E$ have distances determined by \cite{2011MNRAS.417.2500G,2011ApJ...741..110D,2011ApJS..195...14S,2015ApJ...799...29E}, respectively. The remaining sources are assumed to be associated with star formation regions within spiral arms and distances have been determined through the Baysian model of \cite{2016ApJ...823...77R}.  These sources are indicated with a superscript $R$. For those sources where the Baysian model identifies it is unlikely located in a spiral arm (probability $<$ 0.8), the distance is given by their referenced paper (marked by $O$).\\
Column 11: References. Additional information for some sources is contained in the listed references.\\

The 95 GHz methanol masers in the catalog and the maser candidates were detected by a variety of different telescopes and reduced by different groups, using different procedures. This results in variable precision in estimates of sources parameters such as flux density, peak velocity etc. 
Except for the sources detected by \cite{2011ApJS..196....9C,2012ApJS..200....5C,2013ApJ...763....2G} and in this survey, rms noise information was not given for individual sources.
The sources detected by \cite{2006ARep...50..289K} are given in units of antenna temperature, and \cite{2013ApJS..206....9C} are given in units of main beam temperature rather than flux density. In order to facilitate easier comparison of all sources through this catalog, we have converted the antenna temperatures into flux density using the conversion factor given in the original publications. 
The sources detected by \cite{2000MNRAS.317..315V,2005MNRAS.359.1498E} have no information on the integrated flux density $S$.
We assume that the spectra can be described by a Gaussian profile and the integrated flux density $S$ can be derived from the flux density with velocity over the line window: 
$S \equiv \int S(v)dv$ = $P \cdot \Delta V \cdot \sqrt{\pi /4 \ln 2} $, where $P$ is the peak flux density and $\Delta V$ is the line width.
In addition, the results reported by \cite{1995AZh....72...22V} do not give an estimate of the error in the integrated flux density $S$.  

In total the catalog contains information on 481 95 GHz methanol masers.  When assigning distance information, we use a parallax measurement where available \citep{2008PASJ...60..961H,2008PASJ...60...37H,2014ApJ...783..130R,2016arXiv161000242X}, these distances are indicated with a $P$. Then we used the kinematic distances on the basis of the galactic rotation curve, with $\Theta_0$ = 240 km s$^{-1}$, $R_0$ = 8.34 kpc, $\frac{d\Theta}{dR}$ = --0.2 km s$^{-1}$ kpc$^{-1}$, $U_\sun$ = 10.7 km s$^{-1}$, $V_\sun$ = 15.6 km s$^{-1}$, $W_\sun$ = 8.9 km s$^{-1}$, $\overline{U_s}$ = 2.9 km s$^{-1}$, $\overline{V_s}$ = --1.6 km s$^{-1}$, (the A5 model) of \cite{2014ApJ...783..130R}.
Many of the sources in the catalog are within the solar circle and hence there is a kinematic distance ambiguity (KDA) from the galactic rotation model. 
To resolve the KDA, there are three main methods: coincidence with Infrared Dark Clouds (IRDCs), coincidence with H{\sc i} Self Absorption (HISA) and consistency with known kinematic structures in the Galaxy. IRDC are regions that appear dark against the diffuse mid-infrared background \citep[e.g.][]{1998ApJ...494L.199E,2006ApJ...639..227S}, and because they are in front of the majority of the Galactic infrared emission, they are assumed to be at the near kinematic distance. Spatial coincidence between a BGPS source and an IRDC indicates the BGPS source is on the near side of the Galaxy, while BGPS sources without an associated IRDC are assumed to be at the far kinematic distance. The HISA method is similar to the IRDC method, and occurs when the cold neutral medium on the near side absorbs the warmer diffuse background H{\sc i} emission at the same velocity \citep[e.g.][]{2006MNRAS.366.1096B,2009ApJ...690..706A}.  We have utilized literature observations of molecular lines (e.g., NH$_3$, HCO$^+$, N$_2$H$^+$) towards BGPS sources reported by \cite{2011ApJ...741..110D} (marked by $D$) and \cite{2011ApJS..195...14S} (marked by $S$). The former were able to resolve the KDA for 454 BGPS sources using both the IRDC and HISA methods, while the latter identified 192 BGPS sources at the near distance using a combination of observed maser parallax, IRDCs and known kinematic structures in the Galaxy. \citet{2015ApJ...799...29E} (marked by $E$) used Bayesian distance probability density functions to estimate distances for 138 BGPS sources from our sample, resulting in 75 objects with a well-constrained KDA resolution: 40 at the near kinematic distance, 2 at the tangent point, and 23 at the far kinematic distance and 10 in the outer Galaxy. The work of \cite{2015ApJ...799...29E} is based on kinematic distances and utilizes the Galactic Ring Survey $^{13}$CO(1--0) data to morphologically extract velocities for BGPS sources.  For molecular clouds associated with objects with independently established distances (i.e. masers with trigonometric parallaxes and H{\sc ii} regions) they use prior distance probabilities to resolve the KDA for those BGPS sources. Single dish observations \citep[e.g.][]{1994MNRAS.268..464S} have shown that class I and class II methanol masers are often associated, and while high spatial resolution observations \citep[e.g.][]{2009ApJ...702.1615C} show that these two types masers are not co-spatial on arcsecond scales, they are usually driven by the same young stellar object. Due to this close association we also cross-matched for HISA observations towards our BGPS targets by comparing the coordinates with the catalog of \cite{2011MNRAS.417.2500G}  for 518 class II methanol masers (marked by $G$). 
For the sources where we cannot resolve the distance ambiguity we have assigned the near distance, since maser emission is harder to detect at larger distances which makes the near distance more likely. It is important to note that while IRDC or HISA can help to resolve distance ambiguity it is not always reliable. For example, for sources with IRDC associations it is generally assumed that they are at the near distance, while those without an IRDC are assumed to be at the far distance. However, it is not possible to identify an IRDC if it is located in a region without a bright 8 $\mu$m background. Hence, for sources without an identified IRDC, the distances are likely to be assigned the far kinematic distance, but for these sources there is less certainty in resolving the distance ambiguity. For some sources (particularly those with low LSR velocities located close to the direction of the galactic center or anti-center), rotation models do not give a reliable distance estimate.  For some of these we have used the distances assigned in the original papers, for the remainder we have assumed that they are within spiral arms, and estimated a distance from the Bayesian Distance Calculator \citep[marked by $R$,][]{2016ApJ...823...77R}. Note that the Bayesian Distance calculation shows the probability of the masers being located in a spiral arm is greater than 0.8, for those sources where the Baysian model identifies the probability of the masers being located in a spiral arm is less than 0.8, the distance is given by their referenced paper (marked by $O$). For maser candidate L1157B1, associated with a low-mass star formation region we use a distance of 250 pc, given by \cite{2007ApJ...670L.131L}.

\begin{figure}
\plottwo{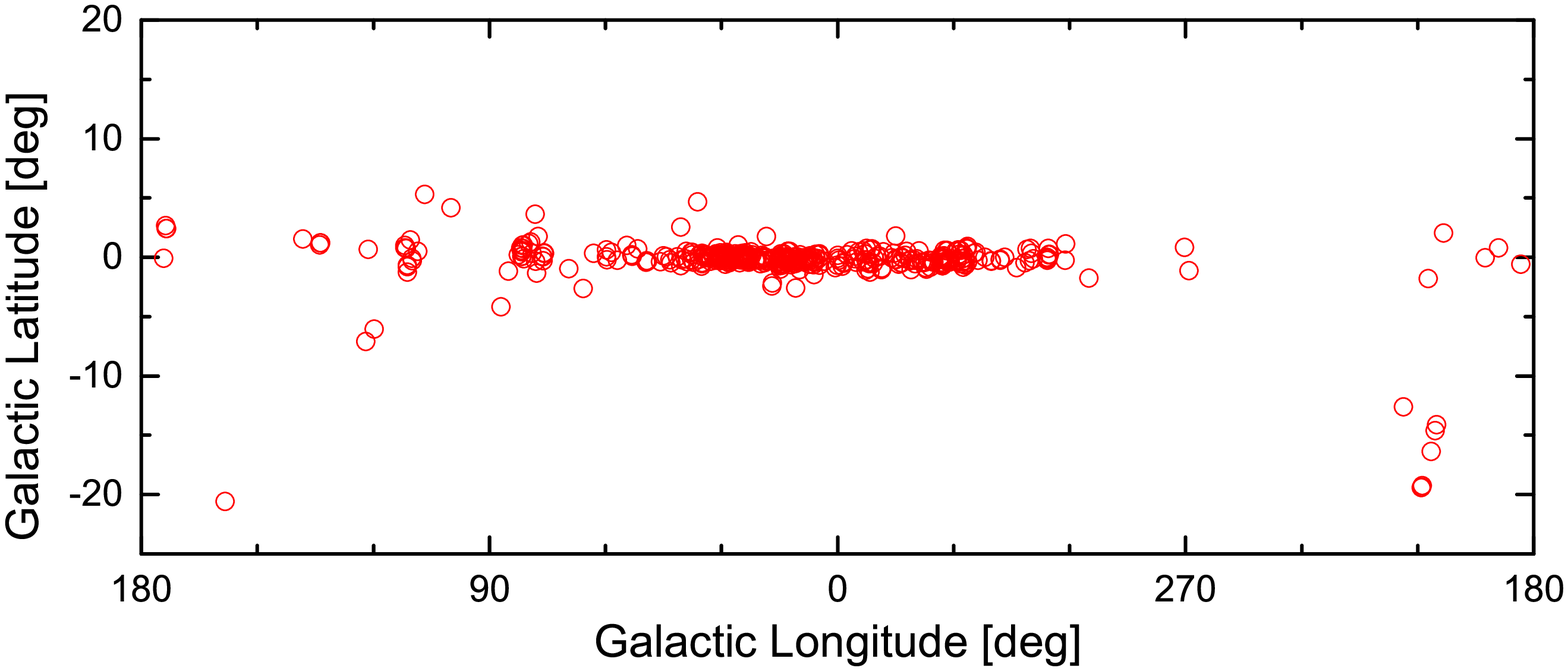}{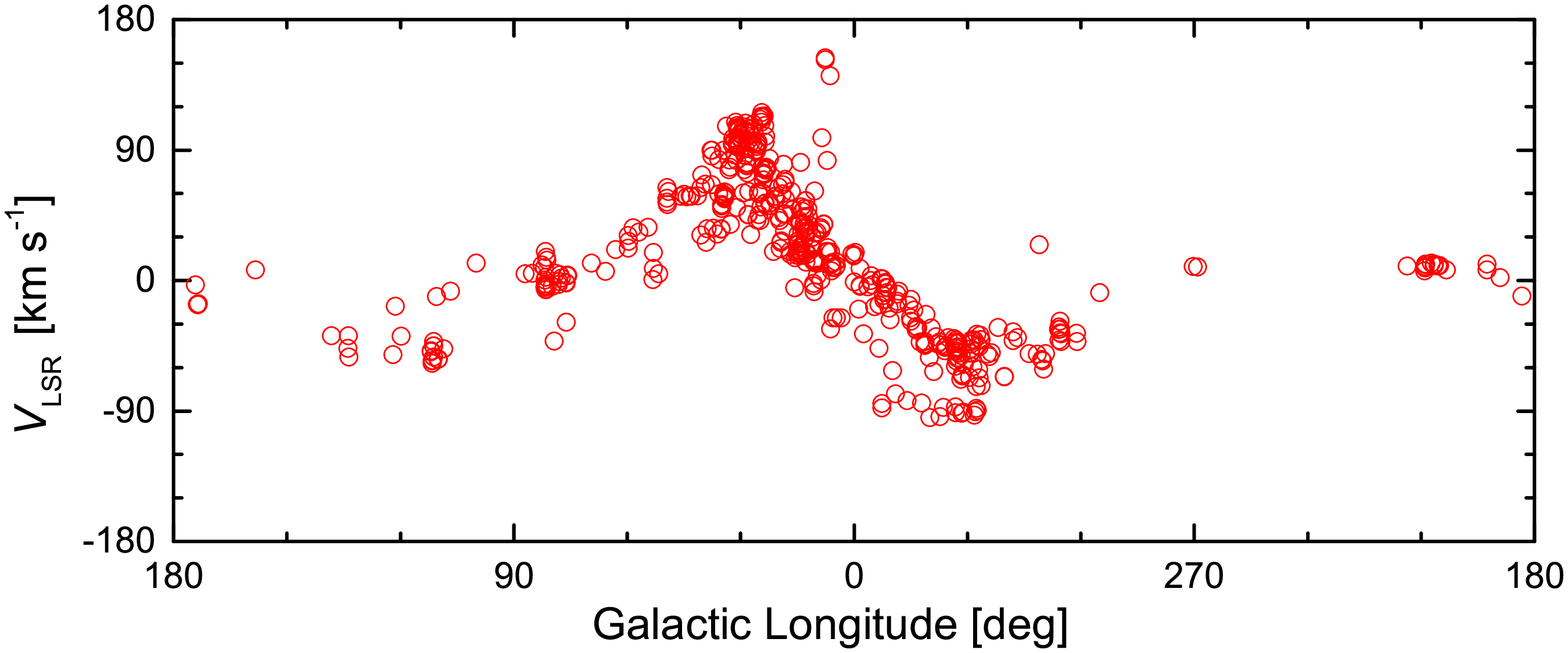}
\plottwo{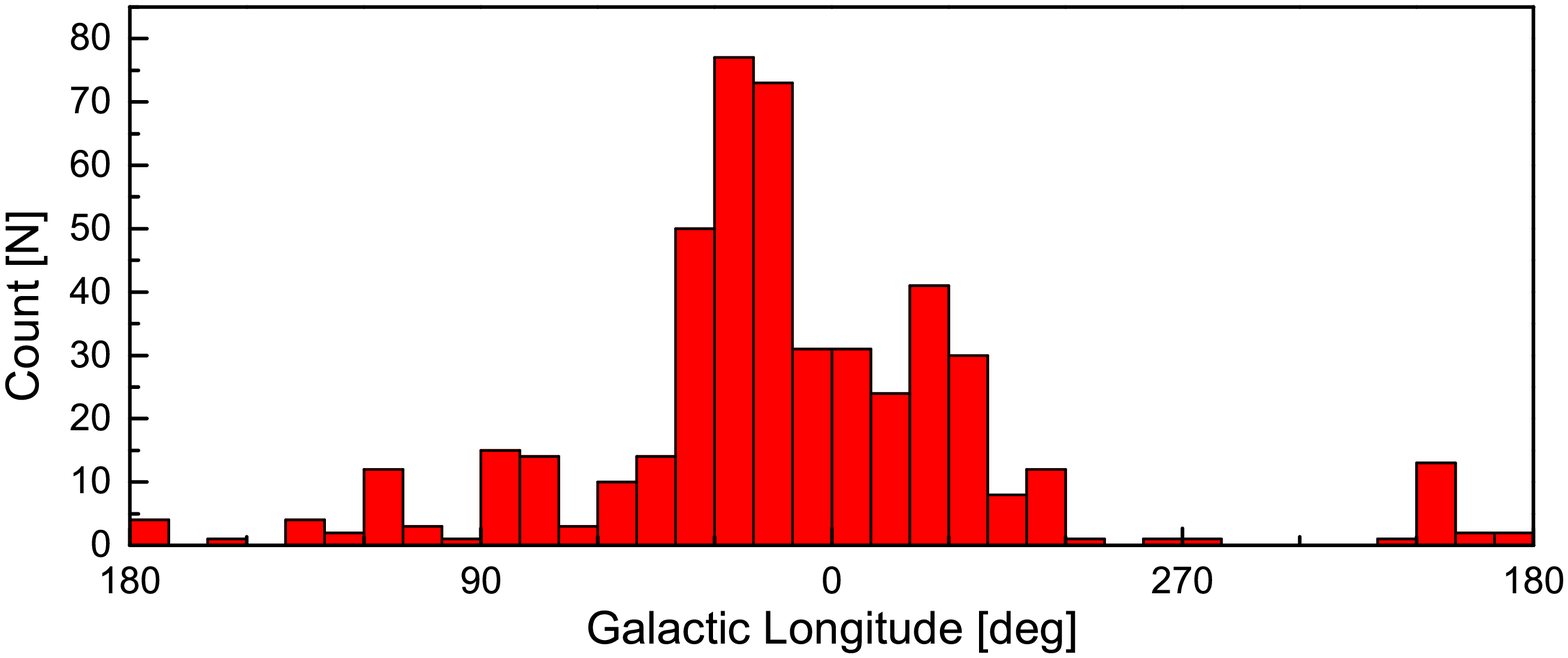}{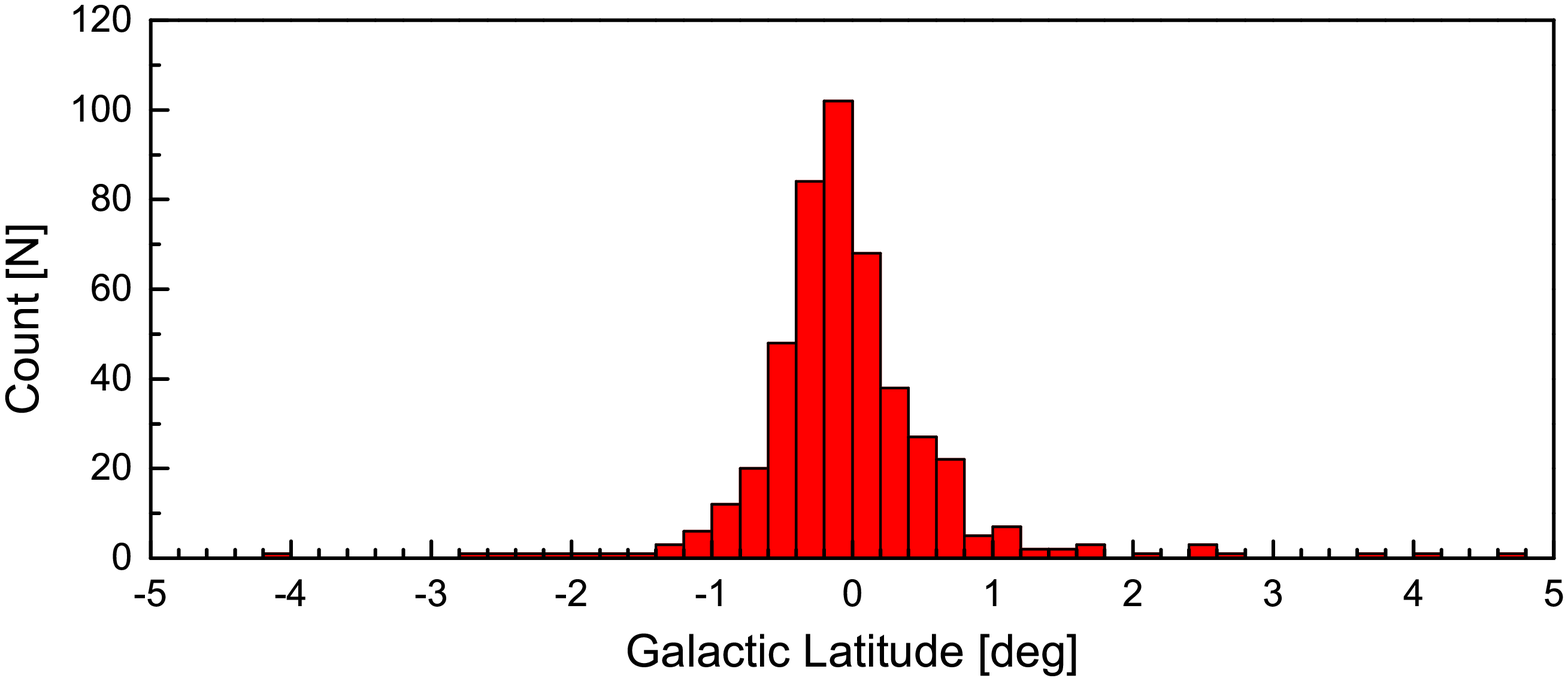}
\caption{\textit{(Upper Left)} Distribution of the 95 GHz methanol masers in the Milky Way. \textit{(Upper Right)} Distribution of the line-of-sight maser peak velocity versus galactic longitude. \textit{(Bottom)} Source counts versus galactic longitude and galactic latitude.\label{fig:3}} 
\end{figure}

The detected 95 GHz methanol masers show a non-random distribution in the Milky Way (See Figure \ref{fig:3}) and Gaussian fitting of the distribution across the galactic plane shows a FWHM of 0.79$^\circ$, indicating that they are concentrated in a central ring of $l$= $\pm$40$^\circ$ with a width of about 1.6$^\circ$ in galactic longitude. Figure \ref{fig:3} also shows some sources at high latitude, seemly offset from the galactic plane. Apart from some high-mass star formation regions with a high latitude but at a near distance (such as DR21, W75N), the other sources which show a significant offset from the Galactic plane (such as NGC1333, IC1396N, HH25MMS) are nearby low-mass star formation regions published in \cite{2013ApJ...763....2G} targeting molecular outflows in low-mass sources.

\section{Summary} \label{sec:summary}

A systematic survey of 95 GHz methanol masers towards 1020 BGPS sources has been made using the PMO 13.7 m radio telescope. 
We detected two hundred and five 95 GHz methanol masers of which 144 are new discoveries, yielding a detection rate of 20\%.
We have compared the properties of HCO$^+$ emission between BGPS sources with or without an associated class I methanol maser, the maser-detected sources show higher values in HCO$^+$ emission suggesting the regions which are hotter and where the star formation activity is stronger are more likely to generate masers. We also compared the velocity difference between the 95 GHz methanol maser peak and that of the corresponding HCO$^+$ thermal emission and found the velocity difference to be less than 1 km s$^{-1}$ in most sources.
Combining all the four hundred and eighty-one 95 GHz methanol masers have been detected so far, we have compiled a catalog to facilitate statistical investigations and future searches. We have also considered thirty-seven 95 GHz methanol emission with broad line width as maser candidates.

Searches for masers using data from infrared and millimeter point source catalogs have been widely used and demonstrated to be very effective for finding new sources.  
However, in many cases they are relatively inefficient, with only a relatively small percentage of the target sources producing maser detections.  
Statistical classification techniques provide a more sophisticated method for predicting potential maser-associated sources than simple color- or flux density-based criteria.  
We find that a variety of different statistical classification models provide comparable levels of accuracy and random forests provides the highest sensitivity. In addition, the detection rate for masers increases with the number of independent models which predict maser emission in that source. BGPS sources which are predicted to have an associated maser by 3 or 4 of the classification techniques by \cite{2016PASA...33...15M} are worth preferential observations in future class I methanol maser searches.\\

$ Acknowledgements $ We thank the anonymous referee for helpful comments that have improved this paper. We are grateful to the staff of the PMO observatory for their assistance during the observations. We also thank Xinyu Du and Yan Gong for helpful suggestions and assistance with English expression. 
This work was supported by the National Science Foundation of China (Grant Numbers: 11673066, 11273043, 11590781 and 11233007) and the Key Laboratory for Radio Astronomy.

\clearpage

\clearpage
\begin{deluxetable}{lccccccccccccc}
\tabletypesize{\scriptsize}
\tablecolumns{11}
\tablewidth{0pt}
\tabcolsep=3pt
\tablecaption{Parameters of 481 All Detected 95 GHz Methanol Masers. \label{tab:MCat}}
\tablehead{
\colhead{Source}   & \colhead{R.A}  & \colhead{Dec} & \colhead{$V_{\rm lsr}$}  & \colhead{$\Delta V$} & \colhead{$P$} & \colhead{$S$}  & \colhead{$S_{\rm int}$} &\colhead{rms} & \colhead{dis}  & \colhead{ref}\\
\colhead{} & \colhead{(J2000)} & \colhead{(J2000)} & 
\colhead{(km s$^{-1}$)}  & \colhead{(km s$^{-1}$)}  & \colhead{(Jy)}  & \colhead{(Jy km s$^{-1}$)}  & \colhead{(Jy km s$^{-1}$)} &\colhead{(Jy)} & \colhead{(kpc)}  & \colhead{}
}
\startdata
G119.80--6.03 & 00:28:44.8 & 56:42:07 & --38.55(0.14) & 2.20(0.38) & 8.37 & 19.62(3.94) & 45.54 & 1.37 & 2.8 & 9\\
  &   &   & --35.66(0.30) & 2.98(0.71) & 5.27 & 16.70(4.15) &   &   &   & \\
  &   &   & --31.95(0.15) & 2.14(0.36) & 4.05 & 9.23(1.46) &   &   &   & \\
G121.30+0.66 & 00:36:47.2 & 63:29:02 & --17.80(0.20) & 2.74(0.48) & 2.57 & 7.49(1.21) & 8.95 & 0.85 & 0.93$^P$ & 9\\
  &   &   & --17.56(0.06) & 0.41(0.28) & 3.29 & 1.45(0.74) &   &   &   & \\
G122.01--7.07 & 00:44:57.2 & 55:47:18 & --51.10(0.00) & 1.09(0.22) & 1.36 & 1.58(0.31) & 3.49 & 0.34 & 2.17$^P$ & 9\\
  &   &   & --48.50(0.00) & 1.56(0.30) & 1.15 & 1.91(0.37) &   &   &   & \\
BGPS7351$^M$ & 02:25:30.8 & 62:06:18 & --45.06(0.18) & 0.56(0.41) & 1.8 & 1.1(0.7) & 10.2 & 0.80 & 4.2$^E$ & *\\
  &   &   & --42.75(0.15) & 2.28(0.37) & 3.8 & 9.2(1.2) &   &   &   & \\
BGPS7364(W3(3))$^M$ & 02:25:53.4 & 62:04:08 & --38.11(0.07) & 1.40(0.20) & 6.1 & 9.1(0.9) & 9.1 & 0.76 & 1.95$^P$ & 2,*\\
BGPS7380(W3(OH))$^M$ & 02:27:04.2 & 61:52:19 & --49.34(0.13) & 1.22(0.35) & 2.6 & 3.4(0.9) & 30.1 & 0.67 & 1.95$^P$ & 2,*\\
  &   &   & --46.30(0.05) & 2.61(0.14) & 9.6 & 26.7(1.1) &   &   &   & \\
\enddata
\tablecomments{
(This table is available in its entirety in machine-readable form. A portion is shown here for guidance.)}
\end{deluxetable}

\begin{deluxetable}{lccccccccccccc}
\tabletypesize{\scriptsize}
\tablecolumns{11}
\tablewidth{0pt}
\tabcolsep=3pt
\tablecaption{Parameters of 95 GHz Methanol Maser Candidates. \label{tab:PCat}}
\tablehead{
\colhead{Source}   & \colhead{R.A}  & \colhead{Dec} & \colhead{$V_{\rm lsr}$}  & \colhead{$\Delta V$} & \colhead{$P$} & \colhead{$S$}  & \colhead{$S_{\rm int}$} &\colhead{rms} & \colhead{dis}  & \colhead{ref}\\
\colhead{} & \colhead{(J2000)} & \colhead{(J2000)} & 
\colhead{(km s$^{-1}$)}  & \colhead{(km s$^{-1}$)}  & \colhead{(Jy)}  & \colhead{(Jy km s$^{-1}$)}  & \colhead{(Jy km s$^{-1}$)} &\colhead{(Jy)} & \colhead{(kpc)}  & \colhead{}
}
\startdata
G123.07--6.31 & 00:52:23.9 & 56:33:45 & --30.73(0.13) & 5.88(0.36) & 5.43 & 34.00(1.61) & 45.28 & 0.81 & 2.82$^P$ & 9\\
  &   &   & --40.31(0.28) & 4.35(0.57) & 2.43 & 11.27(1.33) &   &   &   & \\
NGC1333IRAS2 & 03:29:00.0 & 31:14:19 & 9.97(0.28) & 8.38(0.85) & 2.75 & 24.50(2.00) & 24.50 &   & 0.24$^P$ & 5\\
BGPS7461(AFGL5180)$^{B,M}$ & 06:08:52.9 & 21:38:17 & 3.02(0.04) & 3.16(0.10) & 12.8 & 43.1(1.1) & 43.1 & 0.66 & 2.10$^P$ & 9,*\\
326.641+0.612 & 15:44:33.2 & --54:05:31 & --39.4(0.1) & 4.3(0.2) & 15.8 & 72.3(3.4) & 72.3 &   & 2.6 & 3\\
351.24+0.67 & 17:20:15.8 & --35:54:58 & --4.5(0.1) & 3.4(0.3) & 4.3 & 15.6(1.4) & 15.6 &   & 1.4 & 3\\
SgrA--A & 17:45:52.0 & --28:59:27 & 42.1(0.2) & 18.3(0.4) & 12.1 & 235.7(5.2) & 235.7 &   & 8.3 & 3\\
SgrB2 & 17:47:20.4 & --28:23:07 & 59.0(1.9) & 14.4(4.2) & 34.1 & 522.7(152.5) & 590.8 &   & 7.9$^P$ & 3\\
  &   &   & 70.5(3.4) & 8.2(9.0) & 7.8 & 68.1(74.7) &   &   &   & \\
BGPS1250$^{B,M}$ & 18:02:06.4 & --23:05:11 &  21.72(0.12)  &  2.18(0.31) & 8.9 & 20.7(2.4)  & 20.7 & 1.6  &  3.4 &  *\\
\enddata
\tablecomments{(This table is available in its entirety in machine-readable form. A portion is shown here for guidance.)}
\end{deluxetable}

\clearpage
\begin{deluxetable}{lcccc}
\tablecolumns{5}
\tabletypesize{\scriptsize}
\tabcolsep=3pt
\tablecaption{The results of cross-validating random forests, logistic regression, and LDA classification and prediction for data from the current observations. \label{tab:classification_results}}
\tablehead{
\colhead{} & \colhead{Random} & \colhead{Logistic} & \colhead{} & \colhead{Normalized} \\
\colhead{} & \colhead{forests} & \colhead{regression} & \colhead{LDA} & \colhead{LDA}
}
\startdata
True pos.    & 140  & 133  &  103  & 128\\
False pos.  & 270  & 171  &  140  & 210\\
Flase neg.  & 62  & 69  &  99  & 74\\
True neg.   & 543  & 642  &  672  & 603\\
Sensitivity  & 69\%  & 66\%  &  51\%  & 63\%\\
Specificity  & 67\%  & 79\%  &  83\%  & 74\%\\
\enddata
\end{deluxetable}

\begin{deluxetable}{lcccccr}
\tablecolumns{7}
\tabletypesize{\scriptsize}
\tabcolsep=5pt
\tablecaption{Summary of All the 95 GHz Methanol Masers Survey \label{tab:sum}}
\tablehead{
\colhead{Paper} & \colhead{\#det./\#obs.} & \colhead{Target} & \colhead{Telescope} & \colhead{Beam size [$\arcsec$]} & \colhead{Velocity resolution [km s$^{-1}$]}  & \colhead{ref} 
}
\startdata
\cite{1994AAS..103..129K}   & 9/11 & star-forming regions & Metashovi--13.7m     & 60       & 0.4           & 1\\
\cite{1995AZh....72...22V}       & 35/     &  --                              & Onsala--20m             & --       & --               & 2\\
\cite{2000MNRAS.317..315V}       & 85/153 & 44 GHz masers  & Mopra--22m              & 52   & 0.236    & 3\\
\cite{2005MNRAS.359.1498E}  &26/62  & 6.6 GHz masers & Mopra--22m              & 52   & 0.12     & 4\\
\cite{2006ARep...50..289K}   & 5/6  & bipolar outflows & Onsala--20m             & 39   &   --            & 5\\
\cite{2010AA...517A..56F} & 11/88 &high-mass star-forming regions  & Nobeyama--45m       & 18   & 0.06        & 6\\
\cite{2011ApJS..196....9C} & 105/192 & EGOs    & Mopra--22m              & 36   & 0.11      & 7\\
\cite{2012ApJS..200....5C}  & 63/214 & GLIMPSE point sources+BGPS  & PMO--13.7m             & 53   & 0.19      & 8\\
\cite{2013ApJ...763....2G}   & 62/288 & outflows   & PMO--13.7m             & 55   & 0.13     & 9\\
\cite{2013ApJS..206....9C}  & 39/55  & EGOs & Mopra--22m              & 36   & 0.11     & 10\\
This Paper   & 207/1020 & BGPS sources   & PMO--13.7m             & 55   & 0.19      & *\\
\enddata
\end{deluxetable}

\clearpage
\begin{deluxetable}{ccccccc}
\tablecolumns{7}
\tablewidth{0pt}
\tabcolsep=10pt
\tabletypesize{\scriptsize}
\tablecaption{Parameters of HCO$^{+}$ emission towards in 212 BGPS sources with an associated 95 GHz methanol emission.\label{tab:HCO}}
\tablehead{
\colhead{}    &&  \multicolumn{4}{c}{HCO$^{+}$} \\
\cline{2-7} \\
\colhead{BGPS ID}  & \colhead{$V_{\rm lsr}$} & \colhead{$\Delta V$} & \colhead{$S$} & \colhead{$P$} & \colhead{$S_{\rm int}$} & \colhead{rms}\\
\colhead{} & \colhead{(km s$^{-1}$)} & \colhead{(km s$^{-1}$)} & \colhead{(Jy km s$^{-1}$)} & \colhead{(Jy)} & \colhead{(Jy km s$^{-1}$)} & \colhead{(Jy)}
}

\startdata
1039 & 3.06(0.28) & 6.02(0.64) & 57.8(5.4) & 9.0 & 57.8 & 2.1\\
1062 & 9.41(0.01) & 1.77(0.02) & 110.1(1.1) & 58.3 & 110.1 & 0.9\\
1114 & --27.61(0.06) & 2.65(0.12) & 60.7(3.0) & 21.5 & 139.9 & 1.0\\
 & --23.65(0.07) & 3.51(0.17) & 79.2(3.3) & 21.2 &  & \\
1116 & 9.44(0.03) & 4.59(0.07) & 219.9(2.7) & 45.1 & 220.0 & 1.3\\
1136 & 8.42(0.03) & 3.37(0.07) & 759.5(13.0) & 211.7 & 812.1 & 7.1\\
 & 15.62(0.15) & 1.75(0.32) & 52.6(9.0) & 28.3 &  & \\
1138 & 10.24(0.03) & 3.83(0.08) & 176.5(3.1) & 43.3 & 176.5 & 1.6\\
\enddata
\tablecomments{Column 1: BGPS ID, ordered in increasing Right Ascension. Columns 2--5: the velocity of the peak $V_{\rm lsr}$, the line width $\Delta V$, the integrated flux density $S$ and the peak flux density $P$ of the HCO$^+$ emission in every source estimated from Gaussian fits. The estimated error in the fit parameters are given in parentheses. Column 6: the total integrated flux density $S_{\rm int}$ of the HCO$^+$ spectrum obtained by adding the integrated flux density of all the Gaussian components together. Column 7: 1$\sigma_{\rm rms}$ noise of the observations.
(This table is available in its entirety in machine-readable form. A portion is shown here for guidance.)}
\end{deluxetable}

\clearpage
\begin{table}
\caption{Parameters and Comparison of HCO$^{+}$ With and Without Methanol Masers.\label{tab:HCO-compare}}
\center
\begin{tabular}{c|c|ccc}
\hline
\hline
\multicolumn{2}{c|}{Parameters of} & Line Width  &Peak Flux Density & Total Integrated Flux Density \\
\multicolumn{2}{c|}{HCO$^+$}& [km s$^{-1}$] &  [Jy] &  [Jy km s$^{-1}$] \\
\hline
\multirow{2}{*}{Mean}
& With maser & 3.59 & 38.4 & 244.4 \\
& Without maser & 3.21 & 26.7 & 129.4\\
\hline
\multirow{2}{*}{Median} 
& With maser & 3.22 & 25.7 & 171.7 \\
& Without maser & 3.03 & 20.0 & 104.9\\
\hline
\multicolumn{2}{c|}{$p$-value in K-S test } & 5.0E--2 & 1.3E--4  & 5.9E--12 \\
\hline
\hline
\end{tabular}
\end{table}

\clearpage
\appendix

\section{Appendix information}

\begin{deluxetable}{lccc}
\tablecolumns{4}
\tablewidth{0pt}
\tabcolsep=16pt
\tabletypesize{\scriptsize}
\tablecaption{807 BGPS Sources for which no 95 GHz methanol emission was detected. \label{tab:807}}
\tablehead{
\colhead{BGPS ID}  & \colhead{R.A} & \colhead{DEC} & \colhead{rms}\\
\colhead{} & \colhead{(J2000)} & \colhead{(J2000)} & \colhead{(Jy)}}
\startdata
1055 & 17:55:49.0 & --24:40:13 & 1.6\\
1100 & 17:59:03.6 & --24:20:47 & 0.9\\
1129$^M$ & 18:00:50.7 & --24:10:17 & 1.8\\
1130$^M$ & 18:00:25.1 & --24:06:41 & 0.9\\
1132 & 17:59:51.9 & --24:01:06 & 0.7\\
1135$^M$ & 18:00:22.0 & --24:03:06 & 0.9\\
1139 & 18:00:07.9 & --24:00:17 & 0.7\\
1140$^M$ & 18:00:44.0 & --24:04:43 & 1.7\\
1142$^M$ & 18:01:08.1 & --24:07:11 & 0.9\\
1167 & 17:59:53.2 & --23:45:22 & 0.4\\
\enddata
\tablecomments{Column 1: the source name. Column 2--3: the equatorial coordinates. Column 4: rms noise.\\
(This table is available in its entirety in machine-readable form. A portion is shown here for guidance.)}
\end{deluxetable}

\begin{deluxetable}{ccccccc}
\tablecolumns{7}
\tablewidth{0pt}
\tabcolsep=10pt
\tabletypesize{\scriptsize}
\tablecaption{Information of HCO$^+$ Without Methanol Masers Emission. \label{tab:807HCO}}
\tablehead{
\colhead{}    &&  \multicolumn{4}{c}{HCO$^{+}$} \\
\cline{2-7} \\
\colhead{BGPS ID}  & \colhead{$V_{\rm lsr}$} & \colhead{$\Delta V$} & \colhead{$P$} & \colhead{$S$} & \colhead{$S_{\rm int}$} & \colhead{rms}\\
\colhead{} & \colhead{(km s$^{-1}$)} & \colhead{(km s$^{-1}$)} & \colhead{(Jy)} & \colhead{(Jy km s$^{-1}$)} & \colhead{(Jy km s$^{-1}$)} & \colhead{(Jy)}}
\startdata
1055 & 9.29(0.15) & 1.56(0.50) & 7.8 & 13.0(2.6) & 80.3 & 1.8\\
  & 14.94(0.08) & 3.45(0.20) & 18.3 & 67.2(3.4) &   &  \\
1129 & 16.04(0.02) & 3.46(0.04) & 88.6 & 325.8(3.9) & 325.8 & 1.8\\
1132 & 7.65(0.09) & 1.90(0.19) & 5.5 & 11.1(1.0) & 47.7 & 0.8\\
  & 17.51(0.02) & 1.72(0.06) & 20.0 & 36.6(1.0) &   &  \\
1135 & 6.54(0.02) & 3.52(0.06) & 49.5 & 185.8(2.7) & 216.8 & 0.9\\
  & 17.40(0.14) & 3.47(0.29) & 8.4 & 31.1(2.5) &   &  \\
1139 & 9.51(0.04) & 3.55(0.10) & 14.7 & 55.4(1.3) & 55.4 & 0.7\\
1140 & --4.77(0.13) & 1.33(0.29) & 13.2 & 18.8(3.5) & 201.5 & 3.2\\
  & 7.60(0.06) & 3.56(0.15) & 48.3 & 182.7(6.1) &   &  \\
\enddata
\tablecomments{Column 1: the source name, arranged by increasing BGPS ID. Column 2--5: the velocity of the peak $V_{\rm lsr}$, the line width $\Delta V$, the integrated flux density $S$ and the peak flux density $P$ of the HCO$^+$ emission in every source estimated by Gaussian fits. The fitting error is given in the parentheses. Column 6: the total integrated flux density $S_{\rm int}$ of the HCO$^+$ spectrum obtained by adding the integrated flux density of all the Gaussian components together. Column 7: 1$\sigma_{\rm rms}$ noise of the observations. \\
(This table is available in its entirety in machine-readable form. A portion is shown here for guidance.)}
\end{deluxetable}



\listofchanges

\end{document}